\documentclass[submission,copyright,creativecommons]{eptcs}
\providecommand{\event}{GCM 2024} 

\usepackage{orcidlink}
\usepackage{iftex}

  \title{Modelling Privacy Compliance in Cross-border Data Transfers with Bigraphs}

\author{Ebtihal Althubiti\orcidlink{0009-0005-6218-5121}
\institute{Northern Border University, Arar, Saudi Arabia }
\email{ebtihal.althubiti@nbu.edu.sa}
\institute{ University of Glasgow, Glasgow, United Kingdom }
\email{e.althubiti.1@research.gla.ac.uk}
\and
Michele Sevegnani\orcidlink{0000-0001-6773-9481}
\institute{ University of Glasgow, Glasgow, United Kingdom}
\email{michele.sevegnani@glasgow.ac.uk}
}
\def\titlerunning{Modelling Privacy Compliance in Cross-border Data Transfers with Bigraphs}
\def\authorrunning{E. Althubiti \& M. Sevegnani}

\usepackage{rotating}
\usepackage[utf8]{inputenc}
\usepackage{graphicx}
\usepackage{url}
\graphicspath{ {./images/} }
\usepackage{amsmath, amssymb}
\usepackage{multirow}
\usepackage{enumitem}
\usepackage{array}
\usepackage{caption}
\usepackage{subcaption}
\usepackage{tabularx}
\usepackage{todonotes}
\usepackage{graphics}
\usepackage{booktabs}
\usepackage[capitalise]{cleveref}
\usepackage{xcolor}
\usepackage{listings}

\usetikzlibrary{fit}
\usetikzlibrary{tikzmark}
\usetikzlibrary{positioning}
\usetikzlibrary{calc}
\usetikzlibrary{shapes}
\usetikzlibrary{shapes.geometric}
\usetikzlibrary{shapes.multipart}
\usetikzlibrary{shapes.gates.logic.US}
\usetikzlibrary{arrows}
\usetikzlibrary{arrows.meta}
\usetikzlibrary{backgrounds,scopes}
\tikzset{
  big edge/.style={
    green,
    thick,
  },
  big edgep/.style={
    big edge,
    -{Circle[fill=black,black,width=2,length=2,sep=-1]}
  },
  big pedge/.style={
    big edge,
    {Circle[fill=black,black,width=2,length=2,sep=-1]}-
  },
  big pedgep/.style={
    big edge,
    {Circle[fill=black,black,width=2,length=2,sep=-1]}-{Circle[fill=black,black,width=2,length=2,sep=-1]}
  },
  big edgec/.style={
    big edge,
    -{Bar[fill=green,green,width=4,length=0,sep=0]}
  },
  big pedgec/.style={
    big edge,
    {Circle[fill=black,black,width=2,length=2,sep=-1]}-{Bar[fill=black,black,width=4,length=0,sep=0]}
  },
  big region/.style={
    draw,
    rectangle,
    rounded corners=1.5,
    dashed,
    dash pattern=on 1pt off 1pt,
    thin,
    gray,
  },
  big site/.style={
    big region,
    fill=gray!60,
    text=black,
  },
  big react/.style={
    black,
    thick,
    -stealth,
    line width=3,
    shorten <=3,
    shorten >=3,
  },
  big react rev/.style={
    black,
    thick,
    stealth-stealth,
    line width=3,
    shorten <=3,
    shorten >=3,
  },
  big inst map/.style={
    thick,
    -stealth,
    blue,
    dashed
  },
  lbl/.style={
    font=\tiny\sf,
    inner sep=1,
  },
  lbl conc/.style={
    font=\tiny,
    inner sep=1,
  }
}
\usepackage{amsmath,amsfonts}

\newcommand{\xreact}[1]{\operatorname{\mathrel{\frac{\raisebox{0.75mm}{\begin{scriptsize}\ensuremath{\hspace*{1mm}\ #1 \hspace*{1mm}}\end{scriptsize}}}{}} \joinrel{\!\!\vartriangleright}}}

\DeclareMathOperator{\rrul}{\mathrel{\frac{\raisebox{0.75mm}{\begin{scriptsize}\ensuremath{\hspace*{1mm}\ \hspace*{1mm}}\end{scriptsize}}}{}} \joinrel{\!\!\blacktriangleright}}

\usepackage{xspace}
\newcommand{\ie}{\emph{i.e.}\@\xspace}
\newcommand{\eg}{\emph{e.g.}\@\xspace}

\usepackage{mathrsfs}
\newcommand{\keyword}[1]{\ensuremath{\mathsf{#1}}}
\newcommand{\rr}[1]{\texttt{#1}}
\newcommand{\pred}[1]{\ensuremath{\mathtt{#1}}}

 \usepackage{balance}

\makeatletter
\newcommand*{\etc}{\@ifnextchar{.}{\emph {etc}}{\textit{etc}.\@\xspace}}
\makeatother
\newcommand{\etal}{\emph{et al.}\@\xspace}

\begin{document}
\maketitle

\begin{abstract}
Advancements in information technology have led to the sharing of users’ data across borders, raising privacy concerns, particularly when destination countries lack adequate protection measures. Regulations like the European General Data Protection Regulation (GDPR) govern international data transfers, imposing significant fines on companies failing to comply. To achieve compliance, we propose a privacy framework based on Milner's Bigraphical Reactive Systems (BRSs), a formalism modelling spatial and non-spatial relationships between entities. BRSs evolve over time via user-specified rewriting rules, defined algebraically and diagrammatically. In this paper, we rely on diagrammatic notations, enabling adoption by end-users and privacy experts without formal modelling backgrounds. The framework comprises predefined privacy reaction rules modelling GDPR requirements for international data transfers, properties expressed in Computation Tree Logic (CTL) to automatically verify these requirements with a model checker and sorting schemes to statically ensure models are well-formed. We demonstrate the framework’s applicability by modelling WhatsApp’s privacy policies.
\end{abstract}

\section{Introduction}
Transferring data across various jurisdictions worldwide is one of the requirements for advancing digital systems~\cite{razaghpanah2018apps}. For example, WhatsApp states that they need to share users' data with Meta's data centres and Facebook's branches around the world to improve their services~\cite{whatsapp}. Transferring data to different countries can potentially threaten users' privacy as the receiver country may not use sufficient mechanisms to protect the data, \eg encryption techniques~\cite{casalini2019trade}. 

Several regulations have been imposed to restrict such transfers, for example, the European Union (EU) General Data Protection Regulation (GDPR)~\cite{gdpr}, and the Australian Privacy Principles (APPs)~\cite{australia}.
The main aim of these regulations is to guarantee that the recipient country provides an adequate data protection level compared to the sender country's protection level.

According to the GDPR, there are two main ways to restrict data transfer towards non-EU countries: transferring based on the \textit{adequacy decision} or transferring based on \textit{appropriate safeguards}~\cite{gdprtrans}. The former permits the transfer to specific countries that have an adequate level of data protection as specified by the European Commission. The latter requires safeguards, \eg \textit{Standard Contractual Clauses (SCCs)} or a certification, if the adequacy decision does not cover the receiver country. Besides these two methods, the GDPR identifies some exceptional cases where transferring data is permitted, \eg if the transfer is important for the public interest. We discuss all these ways further in~\cref{sec:privacyReg}.

Organisations must adhere to the above requirements to transfer data from the EU to third countries. Otherwise, they can be at risk of being fined by an EU  authority. For example, Facebook has been fined 1.2 billion euro by the Irish Data Protection Authority (IE DPA) for not adhering to the GDPR requirements by transferring personal data from EU countries to the US\footnote{This is before considering the US as an adequate country.}~\cite{facebook}. Such breaches could occur for many reasons. One is that the regulations are text-based and subject to update, which can create challenges for developers to convert them into technical requirements~\cite{guaman2023automated}. Another reason is that tools used by developers or supervisory authorities to check compliance are not automated, \eg Transfer Impact Assessment (TIA)~\cite{TIA,guaman2023automated}. This motivates the need for an automated approach that enables organisations to ensure their systems comply with the privacy regulations and serve as evidence of their compliance for supervisory authorities. 

Formal methods are mathematical techniques that have been used in various domains to solve these issues thanks to their ability to model systems and provide a rigorous analysis of properties of interest such as security and safety~\cite{woodcock2009formal}. They can also be used as proof of fulfilling these specifications because they enable automated and comprehensive verification processes. 
To the best of our knowledge, they have yet to be used to prove the systems' compliance with the GDPR requirements for cross-border personal data transfers.

In this paper, we  propose a framework based on Milner’s Bigraphs~\cite{milner2009space} to model systems and prove for the first time their compliance to these aspects of privacy regulations.
Bigraphs are a universal formalism for the modelling of interacting systems that evolve in their connectivity and space via rewriting rules called \emph{reaction rules}. 
Compared to other formalisms, reaction rules are user-specified allowing for greater flexibility to model a wide variety of systems.
Bigraphs and reaction rules can be specified algebraically or through an equivalent diagram notation. This feature enables system designers to collaborate with privacy experts, \eg privacy lawyers, who might not be expert in formal modelling and verification. 
Another advantage of using bigraphs is that modelling the flow of data among different jurisdictions requires modelling the spatial relationships between entities, and bigraphs can natively express spatial properties such as containment relation. Additionally, the sorting discipline provides another key advantage by categorising entities and links by means of sorts, ensuring that bigraphs are well-formed and preventing invalid compositions. This adds a layer of structural validation, making the model more robust.

\begin{figure*}[t]
\centering
\includegraphics[width=9cm, height=8.5 cm]{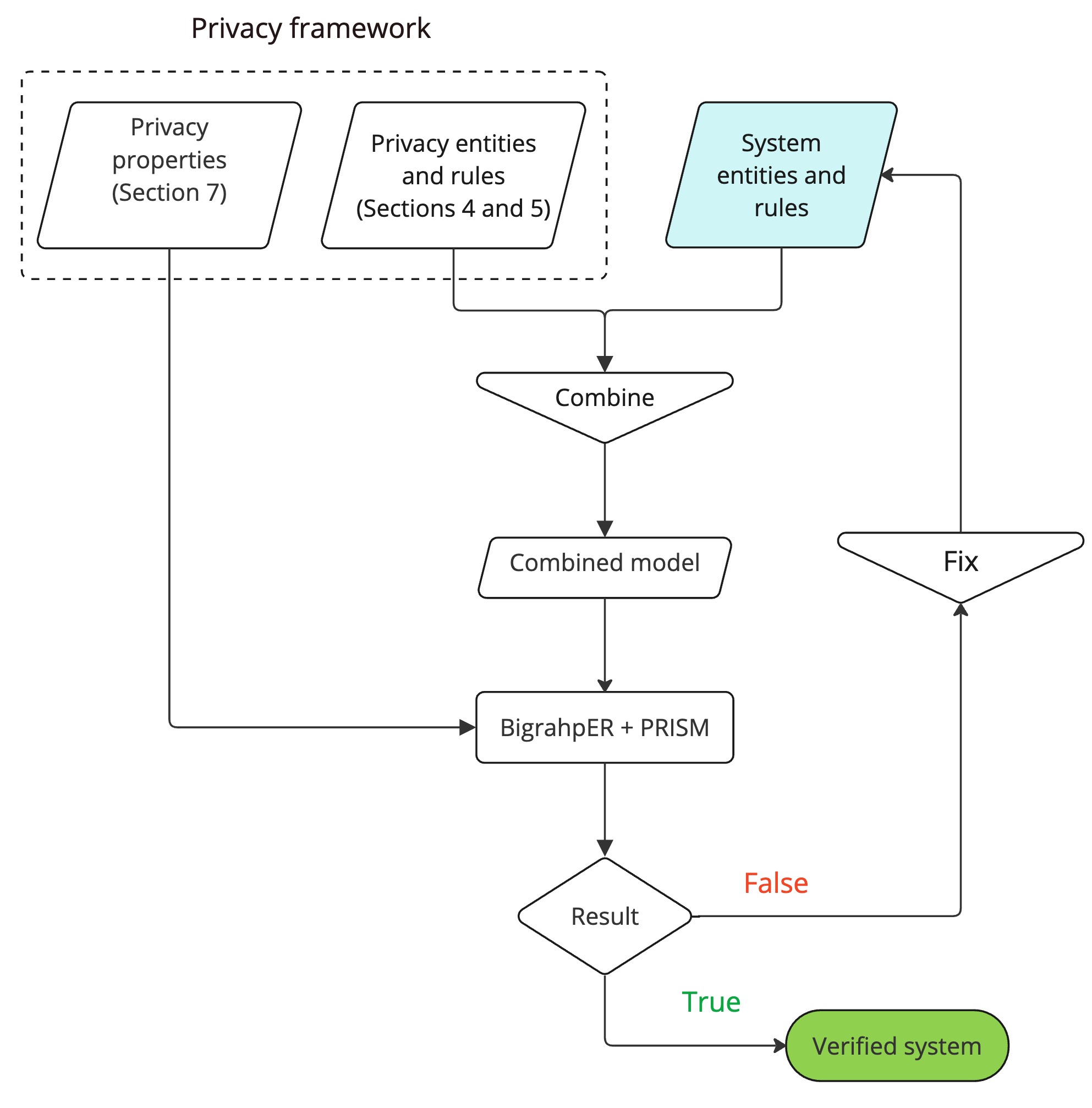}
\caption{Overview of our bigraph-based privacy framework: The teal box represents the system entities and rules that end-users should specify. The process starts by combining the privacy framework with the system's entities and rules (via links) into a unified model. This combined model is then analysed using BigraphER and the PRISM model checker to prove the privacy properties. If the verification result is true, it indicates that the system meets the GDPR requirements for cross-border data transfers. If false, the system model should be fixed to address the identified issues, after which it is reanalysed. This loop continues until the system passes verification.}
\label{fig:framework-overview}
\end{figure*}
Our approach is described in~\cref{fig:framework-overview}. The privacy aspects of our framework are predefined while system-specific aspects have to be specified for each application. The framework also provides a set of privacy properties that should be verified, \eg providing safeguards, the validity of them \etc. These privacy properties are expressed using the Computation Tree Logic (CTL) and can be checked \textit{automatically} using off-the-shelf verification tools such as PRISM~\cite{KNP11}.

We provide the following contributions:
\begin{itemize}
    \item We define a bigraph framework to model the GDPR requirements for cross-border data transfers. The framework captures  \textit{data transfers based on adequacy decisions and appropriate safeguards (Standard Contractual Clauses (SCCs) and certification mechanism)}. To the best of our knowledge, this is the first framework that models the GDPR notion of \textit{international data transfers}.
    \item We apply the framework to an example based on WhatsApp's privacy policies and showcase the capabilities of bigraphs to model complex systems diagrammatically. 
    \item We define the GDPR requirements for international data transfers using CTL and show how they can be proven \textit{formally and automatically} using PRISM.
    \item We define sorting schemes that ensure the model is well-formed by assigning types to entities and links, effectively preventing nonsensical or undesirable model configurations.
    
\end{itemize}
The rest of this paper is structured as follows: \cref{sec:privacyReg} presents the GDPR requirements for international data transfers, while \cref{sec:bigraph_Background} gives an overview of Bigraphical Reactive Systems.
\cref{sec:static_model} and \cref{sec:dynamics} introduce the proposed privacy model, followed by \cref{sec:Integrate}, which demonstrates how we integrate the privacy framework with the specific system. We discuss the formal verification of the model in \cref{sec:verification}, while \cref{sec:sorts} displays the defined sorts of our model. We introduce an example of detecting privacy violations in \cref{sec:violation_ex}. \cref{sec:related_Work} and \cref{sec:discussion} present the related work and the conclusion, respectively.

 \section{Privacy Regulations for Cross-Border Data Transfers}
\label{sec:privacyReg}
Many governments have imposed legislation to regulate transferring data outside their countries or territories. They specify certain conditions to transfer data internationally as such transfers could affect their citizens' privacy or national security (especially if the transferred data is sensitive)~\cite{casalini2019trade}.

For example, the GDPR only allows international transfers in the following three cases:

\begin{enumerate}
    \item \textbf{Adequacy decisions:} this requirement ensures that the recipient country provides a suitable level of data protection. The European Commission has specified a set of countries that provide an adequate level of protection, \eg US, Canada, Japan \etc~\footnote{To view the full list of countries, see \url{https://t.ly/adequacy-decisions}.}.
    \item \textbf{Appropriate safeguards:} if the adequacy decision does not cover the receiver country, then it should provide appropriate safeguards. The GDPR specifies many safeguards. One is adopting \textit{Standard contractual clauses (SCCs)}, a set of predefined standards and rules identified by the European Commission to protect users' data when it is transferred outside EU countries. These rules should be incorporated into the contract between the sender and receiver organisation~\cite{casalini2021mapping}. Another safeguard is using the \textit{certification mechanism}. The certification should meet certain criteria (scheme) related to transparent processing and users' rights, \eg clarifying the storage mechanisms of the data, data subjects rights to access \etc. 
The certification also should be approved by certification bodies that the GDPR specifies, \eg the European Data Protection Board (EDPB)~\cite{certGuidelines}. 
    \item \textbf{Derogations:} the GDPR identifies some special situations in which data transfer is permitted, \eg when transferring data relates to the public interest~\footnote{ For further details, see \url{https://www.edpb.europa.eu/sme-data-protection-guide/international-data-transfers_en}.}.   
\end{enumerate}

 This paper models the transfer based on the adequacy decision and the appropriate safeguards. We focus on modelling the \textit{SCCs and certification mechanism} as the SCCs are commonly used among organisations, and the certification should meet some criteria that need to be checked \cite{gdprtrans}.

\section{Bigraphical Reactive Systems}
\label{sec:bigraph_Background}
An initial bigraph and a set of reaction rules specifying the systems' evolution over time define Bigraphical Reactive Systems (BRSs). Unlike other formalisms, BRSs model systems diagrammatically and algebraically.
In this paper, we only use the diagrammatic representation, but the equivalent algebraic specification is available online~\cite{althubiti_2024_14052899}.

\begin{figure}[t]
     \centering
     \begin{subfigure}[b]{0.35\textwidth}
         \centering
         \includegraphics[ width=\textwidth]{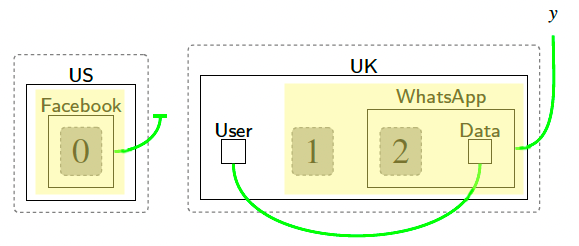}
         \caption{}
         \label{bigraph-b}
     \end{subfigure}
   \quad 
     \begin{subfigure}[b]{0.6\textwidth}
         \centering
         \includegraphics[width=\textwidth]{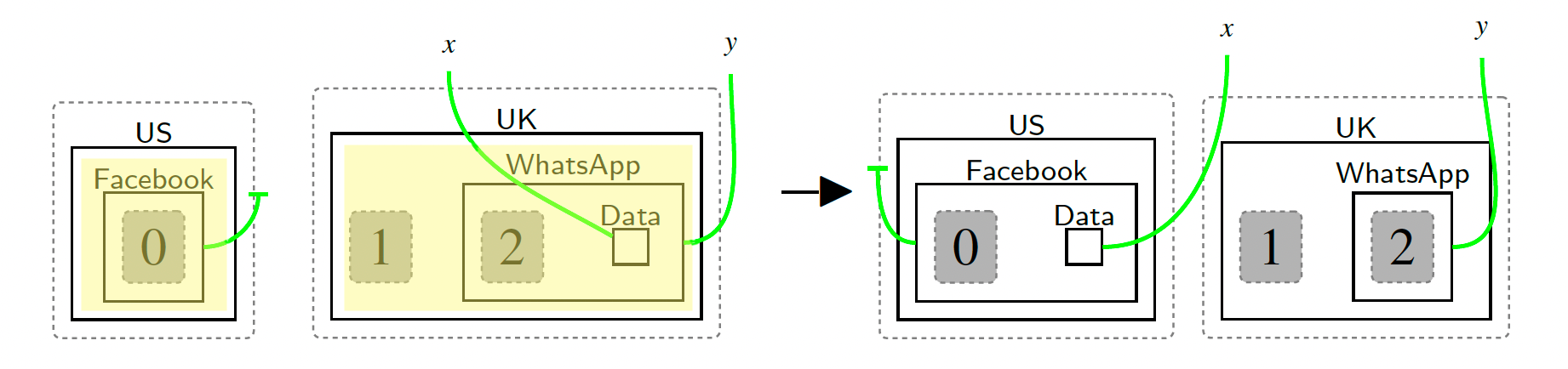}
         \caption{}
         \label{rr:transData}
     \end{subfigure}
     \\
     \begin{subfigure}[b]{0.7\textwidth}
         \centering
         \includegraphics[width=\textwidth]{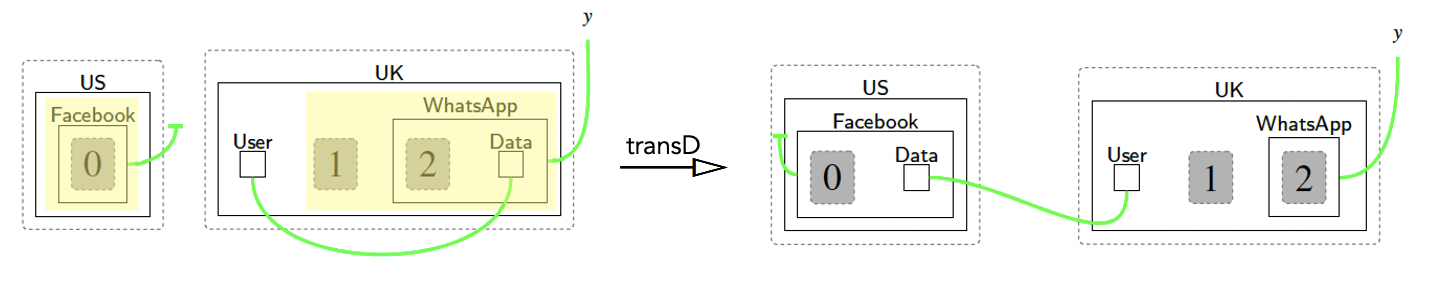}
         \caption{}
         \label{bigraph-Af}
     \end{subfigure}
      \caption{(a) A bigraph representing the initial state of a \keyword{WhatsApp} system that transfers \keyword{Data} to \keyword{Facebook}; (b) Rule \rr{transD} transfers the \keyword{Data} from \keyword{WhatsApp} to \keyword{Facebook}; (c) The result of applying rule \rr{transD} to the initial state, where \keyword{Facebook} acquires the \keyword{Data}.}
        \label{bigraphB}
\end{figure}
An example bigraph is in~\cref{bigraph-b}: a \keyword {WhatsApp} system that transfers \keyword{User}'s \keyword{Data} from the \keyword{UK} to \keyword{Facebook} in the \keyword{US}. Shapes represent entities, \eg \keyword {US}, \keyword {UK}, \etc. 
The shaded rectangles are \emph{sites} that represent components that have been abstracted away as they are irrelevant to the context, \eg site 0 could be a data centre of \keyword {Facebook}.  

The dashed rectangles are called \emph{regions}. Each parallel region can be considered as a modelling perspective that is used to split between different concerns \cite{Benford:2016:LIB:2936307.2882784, DBLP:conf/iceccs/SevegnaniKCM18}, \eg the \keyword{US} is modelled in one perspective and the \keyword{UK} in another.

The green \emph{links} are used to connect entities with each other. Links can be open to indicate the possibility of being linked to other unspecified entities \eg link $y$, or closed to express for example exlusive ownership, \eg the link that connects the {\keyword User} with the {\keyword Data}. Another way to close the links is making them one-to-zero hyperedge, \eg the link that is attached to {\keyword Facebook}. We sometimes use coloured links for readability and visual clarity.
 
Bigraphs can evolve over time by using reaction rules. From now on, we will use rules to indicate reaction rules. Each rule ($L \rrul R$) contains two parts: the left-hand side ($L$) and the right-hand side ($R$). The left-hand side represents the pattern that will be changed, whereas the right-hand side represents the changed pattern. We use the notation $B \xreact{r} B'$ to denote that $B$ rewrites to $B'$ when $r : L \rrul R$ is applied.

For example, rule \rr{transD} in~\cref{rr:transData} models a transfer of the user \keyword {Data} to the \keyword {US}. By applying this rule to bigraph \keyword {B} shown in~\cref{bigraph-b}, the part of the bigraph that matches the left-hand side of the rule (highlighted in yellow) is changed to match the right-hand side \footnote{As this is an illustrative example, we transfer (or optionally duplicate) the data to the \keyword{US} without verifying the GDPR requirements for international data transfers.}. \cref{bigraph-Af} shows the result of applying rule \rr{transD} (\cref{rr:transData}) to bigraph \keyword {B} (\cref{bigraph-b}). The match is performed based on the structure of the bigraphs, not the links' names. This means the links' names are not considered for matching the left-hand side of the rule, \eg changing link $y$ in rule \rr{transD} (\cref{rr:transData}) to $z$ does not affect the application of the rule.

Instantiation maps are a feature of BRS that allows us to copy, swap, or delete sites when the rules are applied. In this paper, we number sites to represent the instantiation maps. For instance, the sites are copied if shown in the left-hand side and the right-hand side of the rule as presented in \cref{rr:transData}. Conversely, they are deleted if they appear in the left-hand side but do not exist in the right-hand side.

We specify and analyse our models using BigraphER~\cite{sevegnani2016bigrapher}, an open-source tool for modelling, rewriting, and visualising bigraphs. BigraphER offers two useful features: (1) enforcement of ordering over the rules via priority classes, \ie each class is a set of rules and any rule in a class with lower priority can only be applied when no rules in any higher priority class can be applied, and (2) parameterized entities and rules, \eg \keyword{Data}($x$) and \rr{transD}($x$), where $x$ can be $\mathtt{Name}$, $\mathtt{ID}$, or $\mathtt{Age}$, to allow sets of entities and rules (each entity or rule using one specific value for $x$).

\section{Modelling Privacy Visually}
\label{sec:static_model}

\begin{figure}[]
    \centering
      \resizebox{0.9 \textwidth}{!}{
       \begin{tikzpicture}[
  ,
_BIG_whatsSystem/.append style = {draw, rounded corners=0.8},
_BIG_whatsLLC/.append style = {draw, rounded corners=0.8},
_BIG_whatsIreland/.append style = {draw, rounded corners=0.8},
_BIG_scheme/.append style = {draw, rounded corners=0.8},
_BIG_sRType/.append style = {draw, rounded corners=0.8},
_BIG_sCCs/.append style = {draw, rounded corners=0.8},
_BIG_proc/.append style = {draw, rounded corners=0.8},
_BIG_p/.append style = {fill , circle ,  inner sep=1.3, black},
_BIG_meta/.append style = {draw, rounded corners=0.8},
_BIG_locations/.append style = {draw, rounded corners=0.8},
_BIG_l/.append style = {draw, rounded corners=0.8},
_BIG_facebookM/.append style = {draw, rounded corners=0.8},
_BIG_facebookD/.append style = {draw, rounded corners=0.8},
_BIG_expiryDate/.append style = {draw, rounded corners=0.8},
_BIG_dC_Ir/.append style = {draw, rounded corners=0.8},
_BIG_contract/.append style = {draw, rounded corners=0.8},
_BIG_cont/.append style = {draw, rounded corners=0.8},
_BIG_cert/.append style = {draw, rounded corners=0.8},
_BIG_c/.append style = {draw, rounded corners=0.8},
_BIG_adeq/.append style = {draw, rounded corners=0.8}
  ]
    
\node[_BIG_p,  label={[inner sep=0.5, name=n16l]north:{\sf\tiny }}] (n16) {};
\node[_BIG_p, right=1.00 of n16, label={[inner sep=0.5, name=n17l]north:{\sf\tiny }}] (n17) {};
\node[_BIG_sCCs, right=1.00 of n17, label={[inner sep=0.5, name=n19l]north:{\sf\tiny SCCs}}] (n19) {};
\node[_BIG_c, right=1.00 of n19, label={[inner sep=0.5, name=n21l]north:{\sf\tiny C($\mathtt{1}$)}}] (n21) {};
\node[_BIG_c, right=0.50 of n21, label={[inner sep=0.5, name=n22l]north:{\sf\tiny C($\mathtt{2}$)}}] (n22) {};
\node[_BIG_c, right=0.50 of n22, label={[inner sep=0.5, name=n23l]north:{\sf\tiny C($\mathtt{3}$)}}] (n23) {};
\node[_BIG_adeq, right=1.00 of n23, label={[inner sep=0.5, name=n24l]north:{\sf\tiny Adeq}}] (n24) {};
\node[_BIG_p, right=1.00 of n24, label={[inner sep=0.5, name=n26l]north:{\sf\tiny }}] (n26) {};
\node[_BIG_p, right=0.50 of n26, label={[inner sep=0.5, name=n27l]north:{\sf\tiny }}] (n27) {};
\node[_BIG_p, right=1.30 of n27, label={[inner sep=0.5, name=n30l]north:{\sf\tiny }}] (n30) {};
\node[_BIG_sCCs, right=0.70 of n30, label={[inner sep=0.5, name=n32l]north:{\sf\tiny SCCs}}] (n32) {};
\node[_BIG_p, right=1.00 of n32, label={[inner sep=0.5, name=n34l]north:{\sf\tiny }}] (n34) {};
\node[_BIG_expiryDate, right=1.00 of n34, label={[inner sep=0.5, name=n36l]north:{\sf\tiny ExpiryDate}}] (n36) {};
\node[_BIG_c, right=0.70 of n36, label={[inner sep=0.5, name=n37l]north:{\sf\tiny C($\mathtt{1}$)}}] (n37) {};
\node[_BIG_c, right=0.50 of n37, label={[inner sep=0.5, name=n38l]north:{\sf\tiny C($\mathtt{2}$)}}] (n38) {};
\node[_BIG_c, right=0.50 of n38, label={[inner sep=0.5, name=n39l]north:{\sf\tiny C($\mathtt{3}$)}}] (n39) {};
\node[_BIG_cont, below=2.1 of n34, label={[inner sep=0.5, name=n1l]north:{\sf\tiny Cont}}] (n1) {};
\node[_BIG_proc, right=0.60 of n1, label={[inner sep=0.5, name=n2l]north:{\sf\tiny Proc}}] (n2) {};
\node[_BIG_cont, right=0.60 of n2, label={[inner sep=0.5, name=n3l]north:{\sf\tiny Cont}}] (n3) {};
\node[_BIG_cont, right=0.60 of n3, label={[inner sep=0.5, name=n4l]north:{\sf\tiny Cont}}] (n4) {};
\node[_BIG_cont, right=0.60 of n4, label={[inner sep=0.5, name=n5l]north:{\sf\tiny Cont}}] (n5) {};
\node[_BIG_proc, right=0.60 of n5, label={[inner sep=0.5, name=n6l]north:{\sf\tiny Proc}}] (n6) {};
\node[_BIG_cert, fit=(n39)(n39l)(n38)(n38l)(n37)(n37l)(n36)(n36l), label={[inner sep=0.5, name=n35l]north:{\sf\tiny Cert}}] (n35) {};
\node[_BIG_contract, fit=(n32)(n32l), label={[inner sep=0.5, name=n31l]north:{\sf\tiny Contract}}] (n31) {};

\node[_BIG_p, right=1.00 of n26, label={[inner sep=0.5, name=e17l]north:{\sf\tiny  }}] (e17) {};

\node[_BIG_l, fit=(n27)(n27l)(n26)(n26l)(e17)(e17l), label={[inner sep=0.5, name=n25l]north:{\sf\tiny L($\mathtt{US}$)}}] (n25) {};
\node[_BIG_scheme, fit=(n23)(n23l)(n22)(n22l)(n21)(n21l), label={[inner sep=0.5, name=n20l]north:{\sf\tiny Scheme}}] (n20) {};
\node[_BIG_contract, fit=(n19)(n19l), label={[inner sep=0.5, name=n18l]north:{\sf\tiny Contract}}] (n18) {};
\node[big site, right=0.6 of n6,] (s3l){};
\node[_BIG_sRType,  fit=(n6)(n6l)(n5)(n5l)(n4)(n4l)(n3)(n3l)(n2)(n2l)(n1)(n1l)(s3l),  label={[inner sep=0.5, name=n0l]north:{\sf\tiny SRType}}] (n0) {};
\node[_BIG_l, fit=(n35)(n35l)(n34)(n34l), label={[inner sep=0.5, name=n33l]north:{\sf\tiny L($\mathtt{Dubai}$)}}] (n33) {};
\node[_BIG_l, fit=(n31)(n31l)(n30)(n30l), label={[inner sep=0.5, name=n29l]north:{\sf\tiny L($\mathtt{Mexico}$)}}] (n29) {};
\node[_BIG_l, fit=(n24)(n24l)(n20)(n20l)(n18)(n18l)(n17)(n17l)(n16)(n16l), label={[inner sep=0.5, name=n15l]north:{\sf\tiny L($\mathtt{Ireland}$)}}] (n15) {};
\node[big site, right=0.8 of n39,] (s1l){};

\node[_BIG_locations, fit=(n33)(n33l)(n29)(n29l)(n25)(n25l)(n15)(n15l)(s1l), label={[inner sep=0.5, name=n14l]north:{\sf\tiny Locations}}] (n14) {};
\node[_BIG_whatsLLC, below =2.1 of n17, label={[inner sep=0.5, name=n8l]north:{\sf\tiny Whats}}] (n8) {};
\node[_BIG_whatsIreland, right=1.30 of n8, label={[inner sep=0.5, name=n9l]north:{\sf\tiny Whats}}] (n9) {};
\node[_BIG_facebookD, right=1.30 of n9, label={[inner sep=0.5, name=n10l]north:{\sf\tiny FB}}] (n10) {};
\node[_BIG_facebookM, right=1.30 of n10, label={[inner sep=0.5, name=n11l]north:{\sf\tiny FB}}] (n11) {};
\node[_BIG_meta, right=1.30 of n11, label={[inner sep=0.5, name=n12l]north:{\sf\tiny Meta}}] (n12) {};
\node[_BIG_dC_Ir, right=1.30 of n12, label={[inner sep=0.5, name=n13l]north:{\sf\tiny DC}}] (n13) {};
\node[big site, right=0.6 of n13,] (s2l){};
{[on background layer]
\node[_BIG_whatsSystem, fit=(n13)(n13l)(n12)(n12l)(n11)(n11l)(n10)(n10l)(n9)(n9l)(n8)(n8l)(s2l), fill=cyan!60!green!13!white, label={[inner sep=0.5, name=n7l]north:{\sf\tiny WhatsSystem}}] (n7) {};}
\node[big region, fit=(n0)(n0l)] (r0) {};
\node[big region, fit=(n7)(n7l)] (r1) {};
\node[big region, fit=(n14)(n14l)] (r2) {};
\coordinate (h0) at ($(n26) + (0.3,-0.5)$);
\draw[big edge,red] (n26) to[out=0,in=-90] (h0);
\draw[big edge,red] (n8) to[out=50,in=-90] (h0);
\draw[big edge,red] (n1) to[out=130,in=-90] (h0);
\coordinate (h1) at ($(n16) + (11.0,-1.7)$);
\draw[big edge,blue] (n16) to[out=-90,in=90] (h1);
\draw[big edge,blue] (n13) to[out=0,in=-90] (h1);
\draw[big edge,blue] (n2) to[out=240,in=-90] (h1);
\coordinate (h2) at ($(n17) + (2.4,-1.0)$);
\draw[big edge,pink!70!black] (n17) to[out=0,in=90] (h2);
\draw[big edge,pink!70!black] (n9) to[out=0,in=-90] (h2);
\draw[big edge,pink!70!black] (n3) to[out=200,in=-90] (h2);
\coordinate (h3) at ($(n27) + (0.3,-0.5)$);
\draw[big edge,black] (n27) to[out=0,in=-90] (h3);
\draw[big edge,black] (n12) to[out=0,in=-90] (h3);
\draw[big edge,black] (n4) to[out=140,in=-90] (h3);
\coordinate (h4) at ($(n34) + (0.3,-0.5)$);
\draw[big edge,yellow!80!black] (n34) to[out=0,in=-90] (h4);
\draw[big edge,yellow!80!black] (n10) to[out=50,in=-90] (h4);
\draw[big edge,yellow!80!black] (n5) to[out=150,in=-90] (h4);
\coordinate (h5) at ($(n30) + (0.3,-0.3)$);
\draw[big edge,green!50!blue!70!yellow] (n30) to[out=0,in=-90] (h5);
\draw[big edge,green!50!blue!70!yellow] (n11) to[out=50,in=-90] (h5);
\draw[big edge,green!50!blue!70!yellow] (n6) to[out=120,in=-90] (h5);
\draw[big edge] (n31) to[out=150,in=20] (n18);
\draw[big edge] (n35) to[out=190,in=-20] (n20);
\draw[big edge] (n24) to[out=28,in=90] (e17);

\end{tikzpicture}      } 
     \caption{A partial initial state for the WhatsApp example. System-specific entities are shaded in teal. The predefined privacy model appears within the uncoloured regions.  }   
     \label{WhatsApp_initialState}
\end{figure}
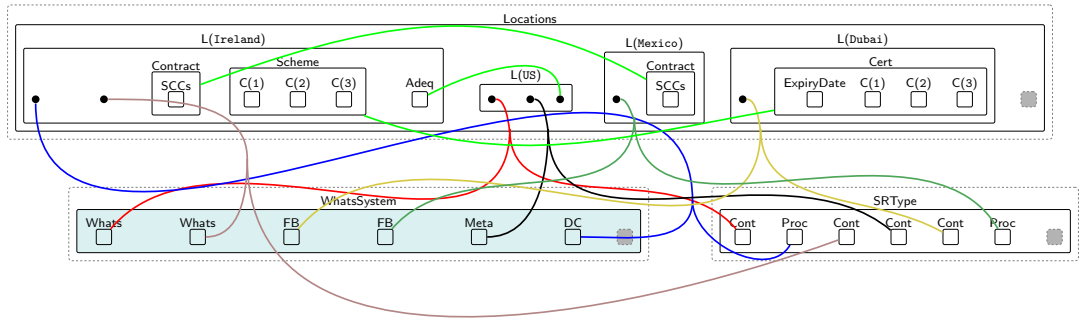

We present our approach by applying it to an example based on WhatsApp's privacy policies. WhatsApp has two \textit{data controllers}~\cite{whatsControllers}, one is located in \keyword{Ireland} to provide services to users in the EU countries, while the other is located in the \keyword{US} to serve users from other countries. 
These two controllers need to share users' data with the parent company \keyword{Meta}, its data centres, and \keyword{Facebook}'s branches worldwide. For space, we present the data centre that is located in \keyword{Ireland} and  
\keyword{Facebook}'s branches in \keyword{Dubai} and \keyword{Mexico}.
A partial model of the example is shown in~\cref{WhatsApp_initialState}, while the full model 
is in~\cite{althubiti_2024_14052899}.
The model consists of three perspectives: \keyword{WhatsSystem}, \keyword{SRType} (for sender and receiver), and \keyword{Locations}. The specific system is modelled in the \keyword{WhatsSystem} perspective. It consists of the data controllers in the US and Ireland (\keyword{Whats}), the Facebook branches in Dubai and Mexico (\keyword{FB}), \keyword{Meta} company in the US, and the data centre in Ireland (\keyword{DC}). These entities are linked to their types in \keyword{SRType} perspective, \eg data controller (\keyword{Cont}) or processor (\keyword{Proc}), and to their locations in \keyword{Locations} perspective as discussed in~\cref{subSec:EntityType_pre} and~\cref{subSec:Locations}.

\subsection{Specifying Sender and Receiver Types }
\label{subSec:EntityType_pre}

\keyword{SRType} perspective is used to specify the types of each system-specific sender and receiver by means of linking, as shown in~\cref{WhatsApp_initialState}.
This perspective can only contain entities of types \textit{data controller} (\keyword{Cont}) and \textit{processor} (\keyword{Proc}). According to the GDPR, the \textit{data controller} determines the privacy policies and the purposes of processing users' data, whereas the processor processes the data based on the controller's instructions~\cite{ControllerResponsib}.
This perspective can easily be extended to support other types, \eg sub-processor, joint controller \etc.
In~\cref{WhatsApp_initialState}, the system's entities that have the controller role (\keyword{Whats}, \keyword{FB} in \keyword{\mathtt{Dubai}}, and \keyword{Meta}) are linked to their own \keyword{Cont}. Similarly, the system's processors (\keyword{DC} and \keyword{FB} in \keyword{\mathtt{Mexico}}) are linked to their own \keyword{Proc}.

\subsection{Specifying Sender and Receiver Locations}
\label{subSec:Locations}
After assigning types to the sender and receiver entities, we need to specify their locations in the \keyword{Locations} perspective.
We do so by linking them to an entity of type \keyword{P} (called \emph{pointer}) nested within a location \keyword{L}($x$). This is shown in~\cref{WhatsApp_initialState} where, for instance,
\keyword{DC} is linked to a \keyword{P} within $\keyword{L}(\mathtt{Ireland})$. This approach allows for a great modelling flexibility as the number of entities in each location does not need to be specified and fixed beforehand. 
Also, by toggling between types \keyword{P}, \keyword{P'} and \keyword{Ps}, it is easy to define rules to check the adequacy decision and the safeguards when the sender and receiver are in different regions as we will explain in~\cref{subSec:Checking_Regions}. 

\begin{table}[t]
    \centering
    \caption{Locations perspective entities.}
    \begin{tabularx}{\columnwidth}{@{}lX@{}}
         \textbf{Entity}     & \textbf{Description} \\ \midrule
         $\keyword{L}(x)$ & Locations as parameterised entities where $x$ is a country's name, \eg $\mathtt{UK}$. \\
         $\keyword{Adeq}$ & Linked to adequate countries. \\
         \keyword{Contract}  &The contract between the sender and the receiver.\\
         \keyword{SCCs}  & The Standard Contractual Clauses. \\
         \keyword{Scheme} & The scheme specified by the GDPR, which the certification should comply with. \\
         \keyword{ExpiryDate}  & The expiry date of the certification. \\
         $\keyword{C}(x)$ & Criteria that should be checked to determine the valid certification. Parameter $x$ is a criterion's identifier, \eg $\mathtt{1}$, $\mathtt{2}$, $\mathtt{Transparency}$, $\mathtt{Right\mbox{-}to\mbox{-}access}$. \\
         \keyword{Cert} & The certification provided by the receiver.  \\
       \keyword{P}, \keyword{P'}, \keyword{Ps} &  Pointers for each entities within a location. Shown as solid black, blue and yellow bullets, respectively. \\
         \end{tabularx}
    \label{tab:LocationsToken}
\end{table}
\cref{tab:LocationsToken} describes the entities of this perspective. The parameterised entity $\keyword{L}(x)$ allows end-users to add new countries, \eg $\keyword{L(\mathtt{Japan})}$. The entity \keyword{SCCs} is nested within \keyword{Contract}. The \keyword{Contract} and \keyword{Cert} are nested within $\keyword{L}(x)$ to indicate that the safeguard provided by the organisation in country $x$ is either the Standard contractual clauses or certification, respectively. Consider the example in~\cref{WhatsApp_initialState}. The entity $\keyword{L}(\mathtt{Mexico})$ contains \keyword{SCCs}, which is nested within the \keyword{Contract}, meaning that the safeguard provided by Facebook in Mexico is the SCCs.
\keyword{Cert} is nested within $\keyword{L}(\mathtt{Dubai})$ to indicate that Facebook's branch in Dubai uses the certification safeguard. We do not allow the case when a company provides both safeguards as this is not included in the GDPR.
Since, $\mathtt{Ireland}$ is the sender country in our example, it includes \keyword{Adeq},  \keyword{Contract}, and \keyword{Scheme}. The scheme contains three criteria, denoted as $\keyword{C}(x)$, where $x \in \{\mathtt{1}, \mathtt{2}, \mathtt{3}\}$.

\section{Checking privacy requirements}
\label{sec:dynamics}
The modelling strategy we have described so far allows to only represent the status of a system at a given time.
To model the temporal evolution of a system, we have also to define a set of reaction rules. In this section, we introduce the reaction rules encoding the processes required to check various privacy requirements. Unlike system-specific rules (see \cref{sec:Integrate}), they can be utilised across systems without changes.

\subsection{Checking Regions}
\label{subSec:Checking_Regions}

To check if data transfers are restricted, we need to check the region of the sender and the receiver. If both are located in the same region, the transfer is considered safe. Otherwise, it is a restricted transfer, and we need to check the adequacy decision and the safeguards.

The mechanism for checking the region is to tag the pointers linked to the sender and recipient \textbf{type}. As these pointers are nested within the entity $\keyword{L}(x)$, tagging them enables us to specify if the sender and receiver are in the same or different regions as explained in~\cref{subSec:SameRegion}.

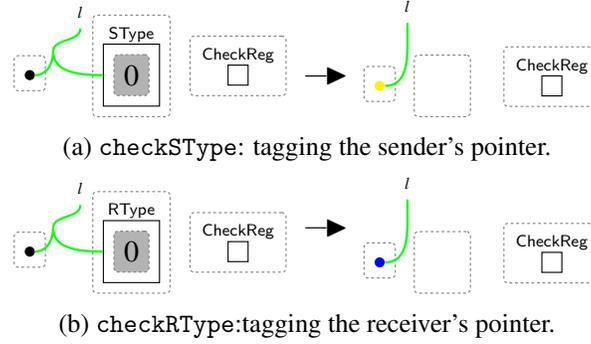
\begin{figure}
     \centering
    
     \begin{subfigure}[b]{0.5\textwidth}
     \resizebox{\textwidth}{!}{
    \begin{tikzpicture}[
    ,
_BIG_p'/.append style = {fill , circle ,  inner sep=1.3, yellow},
_BIG_p/.append style = {fill , circle ,  inner sep=1.3, black},
_BIG_entityType/.append style = {draw},
_BIG_checkReg/.append style = {draw}
    ]
    \begin{scope}[local bounding box=lhs, shift={(0,0)}]
      
\node[_BIG_p,  label={[inner sep=0.5, name=n2l]north:{\sf\tiny }}] (n2) {};
\node[big site, right=1.00 of n2,] (s0l) {0};
\node[_BIG_checkReg, right=1.00 of s0l, label={[inner sep=0.5, name=n0l]north:{\sf\tiny CheckReg}}] (n0) {};
\node[_BIG_entityType, fit=(s0l), label={[inner sep=0.5, name=n1l]north:{\sf\tiny SType}}] (n1) {};
\node[big region, fit=(n0)(n0l)] (r0) {};
\node[big region, fit=(n1)(n1l)] (r1) {};
\node[big region, fit=(n2)(n2l)] (r2) {};
\node[] at ($(r0.north west) + (-1.4,0.3)$) (name_l) {\tiny $l$};
\coordinate (h0) at ($(n2) + (0.3,0.3)$);
\draw[big edge] (n2) to[out=0,in=-90] (h0);
\draw[big edge] (n1) to[out=180,in=-90] (h0);
\draw[big edge] (name_l) to[out=-90,in=90] (h0);

    \end{scope}
    \begin{scope}[local bounding box=rhs, shift={($(lhs.south east) + (1,0)$)}, anchor=south west, scope anchor]
      
\node[_BIG_p',  label={[inner sep=0.5, name=n1l]north:{\sf\tiny }}] (n1) {};
\node[_BIG_checkReg, right=2.00 of n1, label={[inner sep=0.5, name=n0l]north:{\sf\tiny CheckReg}}] (n0) {};
\node[big region, fit=(n0)(n0l)] (r0) {};
\node[big site, opacity=0, right=0.50 of n1] (sfake) {2};
\node[big region, fit=(sfake)] (r3) {};
\node[big region, fit=(n1)(n1l)] (r2) {};
\node[] at ($(r0.north west) + (-1.4,0.3)$) (name_l) {\tiny $l$};
\draw[big edge] (n1) to[out=0,in=-90] (name_l);

    \end{scope}

    \node[xshift=0, yshift=-9] at ($(lhs.east)!0.5!(rhs.west)$) {$\rrul$};
  \end{tikzpicture}      }
     \caption{\rr{checkSType}: tagging the sender's pointer.
}
     \label{rr:checkSType}
     \end{subfigure}
       \qquad\qquad
    \begin{subfigure}[b]{0.5\textwidth}
        \centering
     \resizebox{  \textwidth}{!}{

  \begin{tikzpicture}[
    ,
_BIG_p'/.append style = {fill , circle ,  inner sep=1.3, blue},
_BIG_p/.append style = {fill , circle ,  inner sep=1.3, black},
_BIG_entityType/.append style = {draw},
_BIG_checkReg/.append style = {draw}
    ]
    \begin{scope}[local bounding box=lhs, shift={(0,0)}]
      
\node[_BIG_p,  label={[inner sep=0.5, name=n2l]north:{\sf\tiny }}] (n2) {};
\node[big site, right=1.00 of n2,] (s0l) {0};
\node[_BIG_checkReg, right=1.00 of s0l, label={[inner sep=0.5, name=n0l]north:{\sf\tiny CheckReg}}] (n0) {};
\node[_BIG_entityType, fit=(s0l), label={[inner sep=0.5, name=n1l]north:{\sf\tiny RType}}] (n1) {};
\node[big region, fit=(n0)(n0l)] (r0) {};
\node[big region, fit=(n1)(n1l)] (r1) {};
\node[big region, fit=(n2)(n2l)] (r2) {};
\node[] at ($(r0.north west) + (-1.4,0.3)$) (name_l) {\tiny $l$};
\coordinate (h0) at ($(n2) + (0.3,0.3)$);
\draw[big edge] (n2) to[out=0,in=-90] (h0);
\draw[big edge] (n1) to[out=180,in=-90] (h0);
\draw[big edge] (name_l) to[out=-90,in=90] (h0);

    \end{scope}
    \begin{scope}[local bounding box=rhs, shift={($(lhs.south east) + (1,0)$)}, anchor=south west, scope anchor]
      
\node[_BIG_p',  label={[inner sep=0.5, name=n1l]north:{\sf\tiny }}] (n1) {};
\node[_BIG_checkReg, right=2.00 of n1, label={[inner sep=0.5, name=n0l]north:{\sf\tiny CheckReg}}] (n0) {};
\node[big region, fit=(n0)(n0l)] (r0) {};
\node[big site, opacity=0, right=0.50 of n1] (sfake) {2};
\node[big region, fit=(sfake)] (r3) {};
\node[big region, fit=(n1)(n1l)] (r2) {};
\node[] at ($(r0.north west) + (-1.4,0.3)$) (name_l) {\tiny $l$};
\draw[big edge] (n1) to[out=0,in=-90] (name_l);

    \end{scope}

    \node[xshift=0] at ($(lhs.east)!0.5!(rhs.west)$) {$\rrul$};
  \end{tikzpicture}      }
     \caption{\rr{checkRType}:tagging the receiver's pointer.  }
     \label{rr:checkRType}
     \end{subfigure}
      \caption{ The process of checking the sender's and receiver's regions by tagging the pointers linked to their specified types.}
      \label{rr:checkingReg_All}
\end{figure}

Rule \rr{checkSType} (\cref{rr:checkSType}) specifies the region of the sender, while rule \rr{checkRType} (\cref{rr:checkRType})
specifies the receiver's location.
The entities \keyword{SType}, \keyword{RType}, and \keyword{CheckReg} are generated via system rule(s) in the system perspective, as outlined in \cref{sec:Integrate}. \keyword{SType} and \keyword{RType} represent the types of the sender and receiver, respectively (\eg, site 0 can be either a \keyword{Cont} or \keyword{Proc}), and they are linked to the corresponding location of each. The entity \keyword{CheckReg} is generated by a system rule to trigger the privacy model.

Consider the controller \keyword{Whats}, located in $\keyword{\mathtt{Ireland}}$, which needs to transfer data to the processor \keyword{DC}, also located in $\keyword{\mathtt{Ireland}}$. To model this, we define system rules that specify the type of the sender (\keyword{Whats}) by generating \keyword{SType} around \keyword{Cont} and linking it to the hyper-edge that connects \keyword{Cont} with its pointer in $\keyword{L(\mathtt{Ireland})}$. Similarly, we specify the type for \keyword{DC}, generating \keyword{RType} around \keyword{Proc}, as \keyword{DC} functions as a processor. Additionally, we generate the entity \keyword{CheckReg} to initiate the checking process.

As we generate the \keyword{SType} and \keyword{RType} entities and link them to the locations of the specified types, we apply rule \rr{checkSType} and \rr{checkRType} (\cref{rr:checkingReg_All}) to tag the pointers linked to \keyword{SType} and \keyword{RType}, represented by yellow and blue bullets, respectively.

\subsection{Entities are in the Same Region}
\label{subSec:SameRegion}
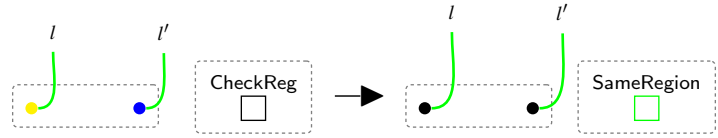
\begin{figure}[]
    \centering
      \resizebox{0.6 \textwidth}{!}{

  \begin{tikzpicture}[
    ,
_BIG_sameRegion/.append style = {draw, green},
_BIG_ps/.append style = {fill , circle ,  inner sep=1.3, yellow},
_BIG_p'/.append style = {fill , circle ,  inner sep=1.3, blue},
_BIG_p/.append style = {fill , circle ,  inner sep=1.3, black},
_BIG_checkReg/.append style = {draw}
    ]
    \begin{scope}[local bounding box=lhs, shift={(0,0)}]
      
\node[_BIG_ps,  label={[inner sep=0.5, name=n1l]north:{\sf\tiny }}] (n1) {};
\node[_BIG_p', right=1.00 of n1, label={[inner sep=0.5, name=n2l]north:{\sf\tiny }}] (n2) {};
\node[_BIG_checkReg, right=1.00 of n2, label={[inner sep=0.5, name=n0l]north:{\sf\tiny CheckReg}}] (n0) {};
\node[big region, fit=(n0)(n0l)] (r0) {};
\node[big region, fit=(n2)(n2l)(n1)(n1l)] (r1) {};
\node[] at ($(r0.north west) + (-1.5,0.3)$) (name_a) {\tiny $l$};
\node[right=0.8 of name_a] (name_a') {\tiny $l'$};
\draw[big edge] (n1) to[out=0,in=-90] (name_a);
\draw[big edge] (n2) to[out=0,in=-90] (name_a');

    \end{scope}
    \begin{scope}[local bounding box=rhs, shift={($(lhs.south east) + (1.0,0)$)}, anchor=south west, scope anchor]
      
\node[_BIG_p,  label={[inner sep=0.5, name=n1l]north:{\sf\tiny }}] (n1) {};
\node[_BIG_p, right=1.00 of n1, label={[inner sep=0.5, name=n2l]north:{\sf\tiny }}] (n2) {};
\node[_BIG_sameRegion, right=1.00 of n2, label={[inner sep=0.5, name=n0l]north:{\sf\tiny SameRegion}}] (n0) {};
\node[big region, fit=(n0)(n0l)] (r0) {};
\node[big region, fit=(n2)(n2l)(n1)(n1l)] (r1) {};
\node[] at ($(r0.north west) + (-1.5,0.3)$) (name_a) {\tiny $l$};
\node[right=0.8 of name_a] (name_a') {\tiny $l'$};
\draw[big edge] (n1) to[out=0,in=-90] (name_a);
\draw[big edge] (n2) to[out=0,in=-90] (name_a');

    \end{scope}

    \node[xshift=0, yshift=-10] at ($(lhs.east)!0.5!(rhs.west)$) {$\rrul$};
  \end{tikzpicture}      } 
     \caption{\rr{sameReg}: the sender and the receiver are in the same region. 
}
\label{rr:sameReg}
\end{figure}

If the tagged pointers (the yellow and blue pointers) are nested within the same region, then the transfer is safe. For instance, the tagged pointers in~\cref{subSec:Checking_Regions} are nested within $\keyword{L(\mathtt{Ireland})}$, so we can verify that the sender and the receiver are both in $\mathtt{Ireland}$. The rule that models the result of checking the regions is \rr{sameReg} (\cref{rr:sameReg}). The entity \keyword{CheckReg} is replaced with \keyword{SameRegion} to indicate the result of checking the region, \ie the sender and the recipient are in the same region, and terminate the checking process.
We also untag the pointers to recheck the regions if needed.

\subsection{Checking Adequacy }
\label{subSec:CheckingAdequacy}

 Suppose the data sender is \keyword{Whats} in $\mathtt{Ireland}$, and the receiver is \keyword{Meta} in the $\mathtt{US}$. By checking their regions as explained in~\cref{subSec:Checking_Regions}, we tag the pointers that are nested within the entities $\keyword{L(\mathtt{Ireland})}$ and $\keyword{L(\mathtt{US})}$, respectively. In such a case, the transfer is restricted because the tagged pointers are nested within different regions, so we must check the adequacy.

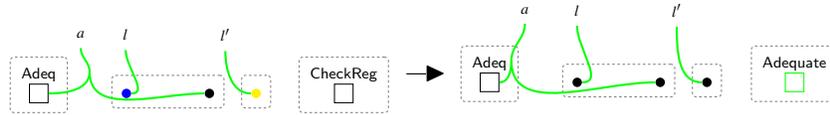
\begin{figure}[]
    \centering
      \resizebox{0.7 \textwidth}{!}{

  \begin{tikzpicture}[
    ,
_BIG_ps/.append style = {fill , circle ,  inner sep=1.3, yellow},
_BIG_p'/.append style = {fill , circle ,  inner sep=1.3, blue},
_BIG_p/.append style = {fill , circle ,  inner sep=1.3, black},
_BIG_checkReg/.append style = {draw},
_BIG_adequate/.append style = {draw, green},
_BIG_adeq/.append style = {draw}
    ]
    \begin{scope}[local bounding box=lhs, shift={(0,0)}]
      
\node[_BIG_adeq,  label={[inner sep=0.5, name=n4l]north:{\sf\tiny Adeq}}] (n4) {};
\node[_BIG_p', right=1.00 of n4, label={[inner sep=0.5, name=n2l]north:{\sf\tiny }}] (n2) {};
\node[_BIG_p, right=1.00 of n2, label={[inner sep=0.5, name=n3l]north:{\sf\tiny  }}] (n3) {};
\node[_BIG_ps, right=0.50 of n3, label={[inner sep=0.5, name=n1l]north:{\sf\tiny }}] (n1) {};
\node[_BIG_checkReg, right=1.00 of n1, label={[inner sep=0.5, name=n0l]north:{\sf\tiny CheckReg}}] (n0) {};
\node[big region, fit=(n0)(n0l)] (r0) {};
\node[big region, fit=(n1)(n1l)] (r1) {};
\node[big region, fit=(n3)(n3l)(n2)(n2l)] (r2) {};
\node[big region, fit=(n4)(n4l)] (r3) {};
\node[] at ($(r0.north west) + (-3.0,0.3)$) (name_a) {\tiny $a$};
\node[right=1.6 of name_a] (name_l) {\tiny $l'$};
\node[left=1 of name_l] (name_l') {\tiny $l$};
\coordinate (h0) at ($(n4) + (0.7,0.3)$);
\draw[big edge] (n4) to[out=0,in=-90] (h0);
\draw[big edge] (n3) to[out=180,in=-90] (h0);
\draw[big edge] (name_a) to[out=-90,in=90] (h0);
\draw[big edge] (n1) to[out=180,in=-90] (name_l);
\draw[big edge] (n2) to[out=0,in=-90] (name_l');

    \end{scope}
    \begin{scope}[local bounding box=rhs, shift={($(lhs.south east) + (1,0)$)}, anchor=south west, scope anchor]
      
\node[_BIG_adeq,  label={[inner sep=0.5, name=n4l]north:{\sf\tiny Adeq}}] (n4) {};
\node[_BIG_p, right=1.00 of n4, label={[inner sep=0.5, name=n2l]north:{\sf\tiny }}] (n2) {};
\node[_BIG_p, right=1.00 of n2, label={[inner sep=0.5, name=n3l]north:{\sf\tiny  }}] (n3) {};
\node[_BIG_p, right=0.50 of n3, label={[inner sep=0.5, name=n1l]north:{\sf\tiny }}] (n1) {};
\node[_BIG_adequate, right=1.00 of n1, label={[inner sep=0.5, name=n0l]north:{\sf\tiny Adequate}}] (n0) {};
\node[big region, fit=(n0)(n0l)] (r0) {};
\node[big region, fit=(n1)(n1l)] (r1) {};
\node[big region, fit=(n3)(n3l)(n2)(n2l)] (r2) {};
\node[big region, fit=(n4)(n4l)] (r3) {};
\node[] at ($(r0.north west) + (-3.3,0.3)$) (name_a) {\tiny $a$};
\node[right=1.7 of name_a] (name_l) {\tiny $l'$};
\node[left=1 of name_l] (name_l') {\tiny $l$};
\coordinate (h0) at ($(n4) + (0.3,0.3)$);
\draw[big edge] (n4) to[out=0,in=-90] (h0);
\draw[big edge] (n3) to[out=180,in=-90] (h0);
\draw[big edge] (name_a) to[out=-90,in=90] (h0);
\draw[big edge] (n1) to[out=180,in=-90] (name_l);
\draw[big edge] (n2) to[out=0,in=-90] (name_l');

    \end{scope}

    \node[xshift=0, yshift=-6] at ($(lhs.east)!0.5!(rhs.west)$) {$\rrul$};
  \end{tikzpicture}      } 
     \caption{\rr{checkingAdeq}: checking the adequacy of the data importer country.}
\label{rr:checkingAdeq}
\end{figure}

 Rule \rr{checkingAdeq} in~\cref{rr:checkingAdeq} checks the adequacy requirement. This rule checks if there is a pointer in the relevant country linked to \keyword{Adeq}. If such a pointer exists, then the country is adequate.
The entity \keyword{Adequate} is generated to show the result of checking the adequacy requirement. The sender's and the receiver's pointers (the yellow and blue bullets, respectively) are untagged to allow reuse of the rule if needed. Importantly, failing to find the match of the left-hand side of this rule means the country is inadequate, so we should start checking the safeguards as shown in~\cref{subSec:CheckingSCC} and \cref{subSec:checkingCert} which have higher priority than rule \rr{checkingAdeq} (\cref{rr:checkingAdeq}).

\subsection{Checking SCCs}
\label{subSec:CheckingSCC} 

\begin{figure}[]
    \centering
      \resizebox{0.5 \textwidth}{!}{

  \begin{tikzpicture}[
    ,
_BIG_p'/.append style = {fill , circle ,  inner sep=1.3, blue},
_BIG_p/.append style = {fill , circle ,  inner sep=1.3, black},
_BIG_invalidContract/.append style = {draw, red},
_BIG_contract/.append style = {draw},
_BIG_checkReg/.append style = {draw}
    ]
    \begin{scope}[local bounding box=lhs, shift={(0,0)}]
      
\node[_BIG_p',  label={[inner sep=0.5, name=n1l]north:{\sf\tiny }}] (n1) {};
\node[_BIG_contract, right=0.50 of n1, label={[inner sep=0.5, name=n2l]north:{\sf\tiny Contract}}] (n2) {};
\node[_BIG_checkReg, right=1.30 of n2, label={[inner sep=0.5, name=n0l]north:{\sf\tiny CheckReg}}] (n0) {};
\node[big region, fit=(n0)(n0l)] (r0) {};
\node[big region, fit=(n2)(n2l)(n1)(n1l)] (r1) {};
\node[] at ($(r0.north west) + (-2.2,0.3)$) (name_a) {\tiny $l$};
\node[right=1.4 of name_a] (name_s) {\tiny $c$};
\draw[big edge] (n1) to[out=180,in=-90] (name_a);
\draw[big edge] (n2) to[out=0,in=-90] (name_s);

    \end{scope}
    \begin{scope}[local bounding box=rhs, shift={($(lhs.south east) + (1,0)$)}, anchor=south west, scope anchor]
      
\node[_BIG_p,  label={[inner sep=0.5, name=n0l]north:{\sf\tiny }}] (n0) {};
\node[_BIG_invalidContract, right=1.00 of n0, label={[inner sep=0.5, name=n1l]north:{\sf\tiny InvalidContract}}] (n1) {};
\node[big region, fit=(n1)(n1l)(n0)(n0l)] (r1) {};
 \node[big site, opacity=0, right=1.30 of n1] (sfake) {2};
\node[big region, fit=(sfake)] (r3) {};
\node[] at ($(r1.north west) + (-0.2,0.3)$) (name_a) {\tiny $l$};
\node[right=1 of name_a] (name_s) {\tiny $c$};
\draw[big edgec] (n1) to[out=-20,in=-90] ($(n1.0) + (0.2,0.1)$);
\draw[big edge] (n0) to[out=180,in=-90] (name_a);

    \end{scope}

    \node[xshift=0, yshift=-6] at ($(lhs.east)!0.5!(rhs.west)$) {$\rrul$};
  \end{tikzpicture}      } 
     \caption{\rr{checkingSCCs}: tagging the contract as invalid if it does not include the SCCs.}
\label{rr:checkingSCCs}
\end{figure}
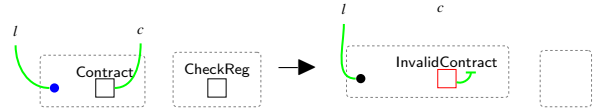

As mentioned previously, we must check the safeguards when rule \rr{checkingAdeq}~(\cref{rr:checkingAdeq}) is not applied. In case the provided safeguard is SCCs, we use rule \rr{checkingSCCs} in~\cref{rr:checkingSCCs} to check whether the receiver country incorporates the \keyword{SCCs} in their \keyword{Contract}.

For example, the \keyword{Contract} of \keyword{FB} in $\mathtt{Mexico}$ includes \keyword{SCCs} and is linked to the entity \keyword{Contract} nested within $\keyword{L(\mathtt{Ireland})}$, as shown in \cref{WhatsApp_initialState}, indicating that $\mathtt{Mexico}$ agrees to adhere to the \keyword{SCCs} specified by the sender, thereby confirming the validity of the \keyword{Contract}.

Conversely, if the \keyword{Contract} in the receiver country does not include \keyword{SCCs}, rule \rr{checkingSCCs} (\cref{rr:checkingSCCs}) is applied to tag the \keyword{Contract} as \keyword{InvalidContract}. This rule also results in the closure of the link connected to the entity \keyword{Contract} in $\mathtt{Ireland}$, \ie link \keyword{c}.

\subsection{Checking Certification}
\label{subSec:checkingCert}

\begin{figure}[]
    \centering
      \resizebox{0.8 \textwidth}{!}{

  \begin{tikzpicture}[
    ,
_BIG_scheme/.append style = {draw},
_BIG_p'/.append style = {fill , circle ,  inner sep=1.3, blue},
_BIG_checkReg/.append style = {draw},
_BIG_cert/.append style = {draw},
_BIG_c'/.append style = {draw, green},
_BIG_c/.append style = {draw}
    ]
    \begin{scope}[local bounding box=lhs, shift={(0,0)}]
      
\node[big site, ] (s1l) {0};
\node[_BIG_c, right=0.50 of s1l, label={[inner sep=0.5, name=n5l]north:{\sf\tiny C(x)}}] (n5) {};
\node[_BIG_p', right=1.0 of n5, label={[inner sep=0.5, name=n1l]north:{\sf\tiny }}] (n1) {};
\node[big site, right=0.50 of n1,] (s0l) {1};
\node[_BIG_c, right=0.50 of s0l, label={[inner sep=0.5, name=n3l]north:{\sf\tiny C(x)}}] (n3) {};
\node[_BIG_checkReg, right=1.30 of n3, label={[inner sep=0.5, name=n0l]north:{\sf\tiny CheckReg}}] (n0) {};
\node[_BIG_scheme, fit=(n5)(n5l)(s1l), label={[inner sep=0.5, name=n4l]north:{\sf\tiny Scheme}}] (n4) {};
\node[_BIG_cert, fit=(n3)(n3l)(s0l), label={[inner sep=0.5, name=n2l]north:{\sf\tiny Cert}}] (n2) {};
\node[big region, fit=(n0)(n0l)] (r0) {};
\node[big region, fit=(n2)(n2l)(n1)(n1l)] (r1) {};
\node[big region, fit=(n4)(n4l)] (r2) {};
\node[] at ($(r0.north west) + (-2.5,0.3)$) (name_a) {\tiny $l$};
\node[left=0.4 of name_a] (name_f) {\tiny $s$};
\draw[big edge] (n1) to[out=50,in=-90] (name_a);
\coordinate (h0) at ($(n4) + (1.2,0.3)$);
\draw[big edge] (n4) to[out=0,in=-90] (h0);
\draw[big edge] (n2) to[out=200,in=-90] (h0);
\draw[big edge] (name_f) to[out=-90,in=90] (h0);

    \end{scope}
    \begin{scope}[local bounding box=rhs, shift={($(lhs.south east) + (1,0)$)}, anchor=south west, scope anchor]
      
\node[big site, ] (s1r) {0};
\node[_BIG_c, right=0.50 of s1r, label={[inner sep=0.5, name=n5l]north:{\sf\tiny C(x)}}] (n5) {};
\node[_BIG_p', right=1.00 of n5, label={[inner sep=0.5, name=n1l]north:{\sf\tiny }}] (n1) {};
\node[big site, right=0.50 of n1,] (s0r) {1};
\node[_BIG_c', right=0.50 of s0r, label={[inner sep=0.5, name=n3l]north:{\sf\tiny C'(x)}}] (n3) {};
\node[_BIG_checkReg, right=1.30 of n3, label={[inner sep=0.5, name=n0l]north:{\sf\tiny CheckReg}}] (n0) {};
\node[_BIG_scheme, fit=(n5)(n5l)(s1r), label={[inner sep=0.5, name=n4l]north:{\sf\tiny Scheme}}] (n4) {};
\node[_BIG_cert, fit=(n3)(n3l)(s0r), label={[inner sep=0.5, name=n2l]north:{\sf\tiny Cert}}] (n2) {};
\node[big region, fit=(n0)(n0l)] (r0) {};
\node[big region, fit=(n2)(n2l)(n1)(n1l)] (r1) {};
\node[big region, fit=(n4)(n4l)] (r2) {};
\node[] at ($(r0.north west) + (-2.5,0.3)$) (name_a) {\tiny $l$};
\node[left=0.4 of name_a] (name_f) {\tiny $s$};
\draw[big edge] (n1) to[out=50,in=-90] (name_a);
\coordinate (h0) at ($(n4) + (1.2,0.3)$);
\draw[big edge] (n4) to[out=0,in=-90] (h0);
\draw[big edge] (n2) to[out=200,in=-90] (h0);
\draw[big edge] (name_f) to[out=-90,in=90] (h0);

    \end{scope}

    \node[xshift=0] at ($(lhs.east)!0.5!(rhs.west)$) {$\rrul$};
  \end{tikzpicture}      } 
     \caption{\rr{tagCriteria(x)}: tagging each criterion that must be satisfied.}   
     \label{rr:tagCriteria}
\end{figure}
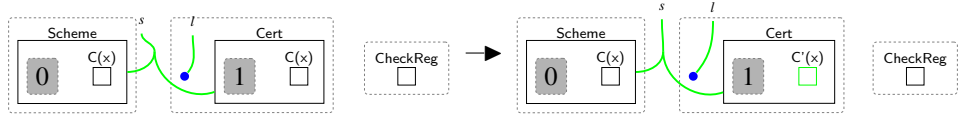

If the provided safeguard is a certification (\keyword{Cert}), it must be checked to prove that it meets the \keyword{Scheme} specified by the GDPR. To do so, we use rule \rr{tagCriteria(x)} shown in~\cref{rr:tagCriteria} to check each criterion~($\mathtt{x}$). 
Suppose that the \keyword{Scheme} has three criteria: $\keyword{C(\mathtt{1})}$, $\keyword{C(\mathtt{2})}$ and $\keyword{C(\mathtt{3})}$. We replace the parameter $\mathtt{x}$ in rule \rr{tagCriteria(x)} (\cref{rr:tagCriteria}) with the number of criterion, \eg \rr{tagCriteria(1)}, \rr{tagCriteria(2)} \etc.

Each rule checks the existence of the criterion in the \keyword{Scheme} and the provided \keyword{Cert}, \eg checking that $\keyword{C(\mathtt{1})}$ is nested within the \keyword{Scheme} and the \keyword{Cert}, by tagging the existing criterion (the green entity). Although the GDPR defines more than three criteria, we model only three as the modeller can extend the model by adding $\mathtt{x}$ number of criteria, \eg $\keyword{C(\mathtt{4})}$, $\keyword{C(\mathtt{5})}$ $\cdots$ $\keyword{C(\mathtt{x})}$.

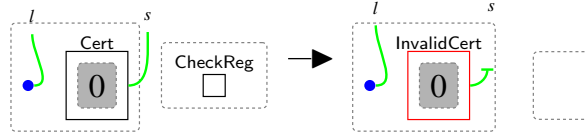
\begin{figure}[]
    \centering
      \resizebox{0.5 \textwidth}{!}{
         \begin{tikzpicture}[
    ,
_BIG_preventTrans/.append style = {draw, thick, circle, red},
_BIG_p'/.append style = {fill , circle ,  inner sep=1.3, blue},
_BIG_invalidCert/.append style = {draw, red},
_BIG_checkReg/.append style = {draw},
_BIG_cert/.append style = {draw}
    ]
    \begin{scope}[local bounding box=lhs, shift={(0,0)}]
      
\node[_BIG_p',  label={[inner sep=0.5, name=n1l]north:{\sf\tiny }}] (n1) {};
\node[big site, right=0.50 of n1,] (s0l) {0};
\node[_BIG_checkReg, right=1.0 of s0l, label={[inner sep=0.5, name=n0l]north:{\sf\tiny CheckReg}}] (n0) {};
\node[_BIG_cert, fit=(s0l), label={[inner sep=0.5, name=n2l]north:{\sf\tiny Cert}}] (n2) {};
\node[big region, fit=(n0)(n0l)] (r0) {};
\node[big region, fit=(n2)(n2l)(n1)(n1l)] (r1) {};
\node[] at ($(r0.north west) + (-1.5, 0.3)$) (name_a) {\tiny $l$};
\node[right=1 of name_a] (name_f) {\tiny $s$};
\draw[big edge] (n1) to[out=0,in=-90] (name_a);
\draw[big edge] (n2) to[out=0,in=-90] (name_f);

    \end{scope}
    \begin{scope}[local bounding box=rhs, shift={($(lhs.south east) + (1,0)$)}, anchor=south west, scope anchor]
      
\node[_BIG_p',  label={[inner sep=0.5, name=n1l]north:{\sf\tiny }}] (n1) {};
\node[big site, right=0.50 of n1,] (s0r) {0};
\node[_BIG_invalidCert, fit=(s0r), label={[inner sep=0.50, name=n2l]north:{\sf\tiny InvalidCert}}] (n2) {};
\node[big site, opacity=0, right=1.0 of s0r] (sfake) {2};
\node[big region, fit=(sfake)] (r0) {};
\node[big region, fit=(n2)(n2l)(n1)(n1l)] (r1) {};
\node[] at ($(r0.north west) + (-2.0,0.3)$) (name_a) {\tiny $l$};
\node[right=1 of name_a] (name_f) {\tiny $s$};
\draw[big edgec] (n2) to[out=0,in=-90] ($(n2.0) + (0.2,0.2)$);
\draw[big edge] (n1) to[out=0,in=-90] (name_a);

    \end{scope}

    \node[xshift=0] at ($(lhs.east)!0.5!(rhs.west)$) {$\rrul$};
  \end{tikzpicture}      } 
     \caption{\rr{tagInvalidCert}: tagging the certification as invalid if at least one criterion is not satisfied.}   
     \label{rr:tagInvalidCert}
\end{figure}

If \textbf{one} of the defined family rules is not applied, the \keyword{Cert} does not meet the criteria. For example, if rule \rr{tagCriteria(1)} is not applied but rule \rr{tagCriteria(2)} is applied, it means the \keyword{Cert} does not fulfil the first criterion ($\keyword{C(\mathtt{1})}$). In such a case, we use rule \rr{tagInvalidCert} (\cref{rr:tagInvalidCert}) to tag the \keyword{Cert} as invalid (\keyword{InvalidCert}) and close link $s$ that is connected to the \keyword{Scheme}. We omit \keyword{CheckReg} to end the checking process. 

\subsection{Result of Checking Certification}
\label{subSec:CheckingExpiry}

\begin{figure}[ht]
    \centering
      \resizebox{0.8 \textwidth}{!}{

  \begin{tikzpicture}[
    ,
_BIG_cuurentDate/.append style = {draw},
_BIG_compliantCert/.append style = {draw},
_BIG_checkReg/.append style = {draw},
_BIG_checkExp/.append style = {draw},
_BIG_cert/.append style = {draw},
_BIG_c'/.append style = {draw, green}
    ]
    \begin{scope}[local bounding box=lhs, shift={(0,0)}]
      
\node[big site, ] (s0l) {0};
\node[_BIG_c', right=0.50 of s0l, label={[inner sep=0.5, name=n2l]north:{\sf\tiny C'(1)}}] (n2) {};
\node[_BIG_c', right=0.5 of n2, label={[inner sep=0.5, name=n3l]north:{\sf\tiny C'(2)}}] (n3) {};
\node[_BIG_c', right=0.5 of n3, label={[inner sep=0.5, name=n4l]north:{\sf\tiny C'(3)}}] (n4) {};
\node[_BIG_checkReg, right=1.40 of n4, label={[inner sep=0.5, name=n0l]north:{\sf\tiny CheckReg}}] (n0) {};
\node[_BIG_cert, fit=(n4)(n4l)(n3)(n3l)(n2)(n2l)(s0l), label={[inner sep=0.5, name=n1l]north:{\sf\tiny Cert}}] (n1) {};
\node[big region, fit=(n0)(n0l)] (r0) {};
\node[big region, fit=(n1)(n1l)] (r1) {};
\node[] at ($(r0.north west) + (-4.0,0.8)$) (name_f) {\tiny $s$};
\draw[big edge] (n1) to[out=180,in=-90] (name_f);

    \end{scope}
    \begin{scope}[local bounding box=rhs, shift={($(lhs.south east) + (1,0)$)}, anchor=south west, scope anchor]
      
\node[big site, ] (s0r) {0};
\node[_BIG_c', right=0.50 of s0r, label={[inner sep=0.5, name=n3l]north:{\sf\tiny C'(1)}}] (n3) {};
\node[_BIG_c', right=0.5 of n3, label={[inner sep=0.5, name=n4l]north:{\sf\tiny C'(2)}}] (n4) {};
\node[_BIG_c', right=0.5 of n4, label={[inner sep=0.5, name=n5l]north:{\sf\tiny C'(3)}}] (n5) {};
\node[_BIG_cuurentDate, right=1.80 of n5, label={[inner sep=0.5, name=n1l]north:{\sf\tiny CurrentDate}}] (n1) {};
\node[_BIG_compliantCert, fit=(n5)(n5l)(n4)(n4l)(n3)(n3l)(s0r), label={[inner sep=0.5, name=n2l]north:{\sf\tiny CompliantCert}}] (n2) {};
\node[_BIG_checkExp, fit=(n1)(n1l), label={[inner sep=0.5, name=n0l]north:{\sf\tiny CheckExp}}] (n0) {};
\node[big region, fit=(n0)(n0l)] (r0) {};
\node[big region, fit=(n2)(n2l)] (r1) {};
\node[] at ($(r0.north west) + (-4.0,0.3)$) (name_f) {\tiny $s$};
\draw[big edge] (n2) to[out=180,in=-90] (name_f);

    \end{scope}

    \node[xshift=0] at ($(lhs.east)!0.5!(rhs.west)$) {$\rrul$};
  \end{tikzpicture}      } 
     \caption{\rr{checkingCertResult}: result of checking the certification and initialising the process of checking its expiry date. }   
     \label{rr:checkingCertResult}
\end{figure}
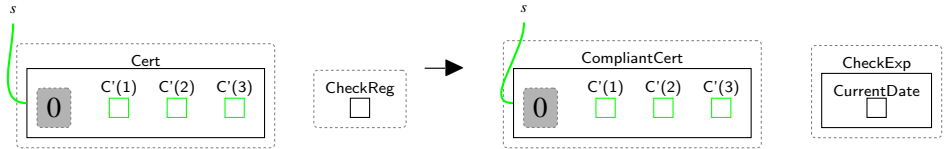

Rule \rr{checkingCertResult} (\cref{rr:checkingCertResult}) shows the result of checking the criteria. If the \keyword{Cert} meets the criteria, \ie $\keyword{C(\mathtt{1})}$, $\keyword{C(\mathtt{2})}$ and $\keyword{C(\mathtt{3})}$ are tagged, the \keyword{Cert} is tagged as \keyword{CompliantCert}.

Based on the GDPR, the certification (\keyword{Cert}) is valid for only \textbf{three years}~\cite{certExpireDate}. This requires checking its validation date as well~\cite{certGuidelines}. Rule~\rr{checkingCertResult} (\cref{rr:checkingCertResult}) initialises the process of checking the expiration by replacing the entity \keyword{CheckReg} with \keyword{CheckExp} and specifying the current date (\keyword{CurrentDate}) to compare it with the expiry date of the \keyword{Cert} as shown in~\cref{subSec:ResultCeckingExpDate}

\subsection{Checking the Expiry Date }
\label{subSec:ResultCeckingExpDate}
\begin{figure}[ht!]
    \centering
      \resizebox{0.6 \textwidth}{!}{

  \begin{tikzpicture}[
    ,
_BIG_cert/.append style = {draw},
_BIG_rejected/.append style = {draw,red},
_BIG_greater/.append style = {draw},
_BIG_preventTrans/.append style = {draw, thick, circle, red},
_BIG_expiryDate/.append style = {draw},
_BIG_cuurentDate/.append style = {draw},
_BIG_compliantCert/.append style = {draw},
_BIG_checkExp/.append style = {draw}
    ]
    \begin{scope}[local bounding box=lhs, shift={(0,0)}]
      
\node[big site, ] (s0l) {0};
\node[_BIG_expiryDate, right=0.80 of s0l, label={[inner sep=0.5, name=n3l]north:{\sf\tiny ExpiryDate}}] (n3) {};
\node[_BIG_cuurentDate, right=1.80 of n3, label={[inner sep=0.5, name=n1l]north:{\sf\tiny CurrentDate}}] (n1) {};
\node[_BIG_compliantCert, fit=(n3)(n3l)(s0l), label={[inner sep=0.5, name=n2l]north:{\sf\tiny CompliantCert}}] (n2) {};
\node[_BIG_checkExp, fit=(n1)(n1l), label={[inner sep=0.5, name=n0l]north:{\sf\tiny CheckExp}}] (n0) {};
\node[big region, fit=(n0)(n0l)] (r0) {};
\node[big region, fit=(n2)(n2l)] (r1) {};
\node[] at ($(r0.north west) + (-0.3,0.3)$) (name_f) {\tiny $s$};
\draw[big edge] (n2) to[out=0,in=-90] (name_f);

    \end{scope}
    \begin{scope}[local bounding box=rhs, shift={($(lhs.south east) + (1,0)$)}, anchor=south west, scope anchor]
      
\node[big site, ] (s0r) {0};
\node[_BIG_expiryDate, right=0.90 of s0r, label={[inner sep=0.5, name=n2l]north:{\sf\tiny ExpiryDate}}] (n2) {};
\node[_BIG_greater, fit=(n2)(n2l), label={[inner sep=0.5, name=n1l]north:{\sf\tiny Greater}}] (n1) {};
\node[_BIG_rejected, fit=(n1)(n1l)(s0r), label={[inner sep=0.5, name=n0l]north:{\sf\tiny InvalidCert}}] (n0) {};
\node[big region, ] (r0) {};
\node[big region, fit=(n0)(n0l)] (r1) {};
\node[big site, opacity=0, right=1.5 of n2] (sfake) {2};
\node[big region, fit=(sfake)] (r3) {};
\node[] at ($(r0.north west) + (2.7,1.3)$) (name_f) {\tiny $s$};
\draw[big edgec] (n0) to[out=0,in=-90] ($(n0.0) + (0.3,0.5)$);

    \end{scope}

    \node[xshift=0] at ($(lhs.east)!0.5!(rhs.west)$) {$\rrul$};
  \end{tikzpicture}      } 
     \caption{\rr{ExpiredDate}: tagging the expired certification. }   
     \label{rr:ExpiredDate}
\end{figure}
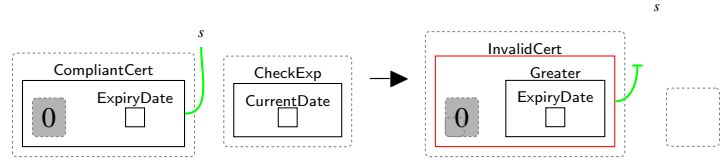

As shown in~\cref{rr:ExpiredDate}, we can check the validation date of the \keyword{CompliantCert} by comparing the current date (\keyword{CurrentDate}) with the expiry date (\keyword{ExpiryDate}). If the \keyword{ExpiryDate} is \textbf{greater than} (\keyword{Greater}) the \keyword{CurrentDate}, it means the \keyword{CompliantCert} is expired and cannot be used as a safeguard. Otherwise, it is valid and can be used to transfer the data~\footnote{ To generalise the model, we do not use mathematical computation to perform the comparison. However, we model the two cases: when the certification is expired and when it is valid. See the full model \cite{althubiti_2024_14052899}.}. 

Rule~\rr{ExpiredDate}~(\cref{rr:ExpiredDate}) tags the \keyword{CompliantCert} as \keyword{InvalidCert} and removes link $s$ connecting the \keyword{CompliantCert} to the \keyword{Scheme} if the \keyword{ExpiryDate} is \keyword{Greater} than the \keyword{CurrentDate}. The entities \keyword{CheckExp} and \keyword{CurrentDate} are omitted to terminate the checking process.

\subsection{Withdrawing Certification}
\label{subSec:WithdCert}
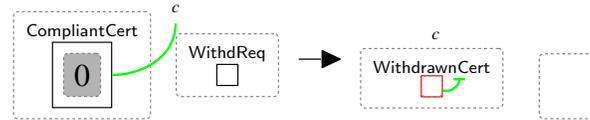
\begin{figure}[]
    \centering
      \resizebox{0.5 \textwidth}{!}{

  \begin{tikzpicture}[
    ,
_BIG_withdrawnCert/.append style = {draw, red},
_BIG_withdReq/.append style = {draw},
_BIG_compliantCert/.append style = {draw}
    ]
    \begin{scope}[local bounding box=lhs, shift={(0,0)}]
      
\node[big site, ] (s0l) {0};
\node[_BIG_withdReq, right=1.40 of s0l, label={[inner sep=0.5, name=n0l]north:{\sf\tiny WithdReq}}] (n0) {};
\node[_BIG_compliantCert, fit=(s0l), label={[inner sep=0.5, name=n1l]north:{\sf\tiny CompliantCert}}] (n1) {};
\node[big region, fit=(n0)(n0l)] (r0) {};
\node[big region, fit=(n1)(n1l)] (r1) {};
\node[] at ($(r0.north west) + (0,0.3)$) (name_f) {\tiny $c$};
\draw[big edge] (n1) to[out=0,in=-90] (name_f);

    \end{scope}
    \begin{scope}[local bounding box=rhs, shift={($(lhs.south east) + (1,0)$)}, anchor=south west, scope anchor]
      
\node[_BIG_withdrawnCert,  label={[inner sep=0.5, name=n0l]north:{\sf\tiny WithdrawnCert}}] (n0) {};
\node[big region, ] (r0) {};
\node[big site, opacity=0, right=1.3 of n0] (sfake) {2};
\node[big region, fit=(sfake)] (r3) {};
\node[big region, fit=(n0)(n0l)] (r1) {};
\node[] at ($(r0.north west) + (0,0.3)$) (name_f) {\tiny $c$};
\draw[big edgec] (n0) to[out=-20,in=-90] ($(n0.0) + (0.2,0.1)$);

    \end{scope}

    \node[xshift=0] at ($(lhs.east)!0.5!(rhs.west)$) {$\rrul$};
  \end{tikzpicture}

      } 
     \caption{\rr{processWithdReq}: processing the withdrawal request for the certification.}   
     \label{rr:processWithdReq}
\end{figure}

Based on the GDPR, the provider of the compliant certification (\keyword{CompliantCert}) has the right to withdraw it~\cite{certGuidelines}. Rule \rr{processWithdReq} (\cref{rr:processWithdReq}) models the provider's withdrawal request by closing link $c$ connecting the \keyword{CompliantCert} to the \keyword{Scheme} and replacing \keyword{CompliantCert} with \keyword{WithdrawnCert}. The entity \keyword{WithdReq} is generated through a system rule (see rule \rr{ReqWithdCert} in \cref{rr:ReqWithdCert}).

\section{Integration of Privacy Models and Specific System}
\label{sec:Integrate}

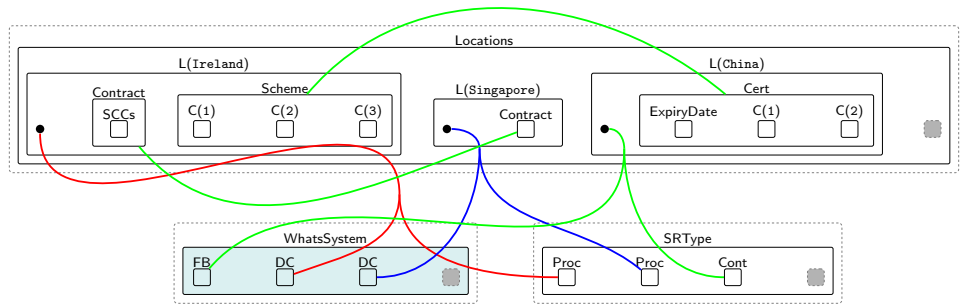
\begin{figure*}[]
    \centering
      \resizebox{0.8 \textwidth}{!}{
       \begin{tikzpicture}[
  ,
_BIG_whatsSystem/.append style = {draw, rounded corners=0.8},
_BIG_scheme/.append style = {draw, rounded corners=0.8},
_BIG_sRType/.append style = {draw, rounded corners=0.8},
_BIG_sCCs/.append style = {draw, rounded corners=0.8},
_BIG_proc/.append style = {draw, rounded corners=0.8},
_BIG_p/.append style = {fill , circle ,  inner sep=1.3, black},
_BIG_locations/.append style = {draw, rounded corners=0.8},
_BIG_l/.append style = {draw, rounded corners=0.8},
_BIG_facebookC/.append style = {draw, rounded corners=0.8},
_BIG_expiryDate/.append style = {draw, rounded corners=0.8},
_BIG_dC_Si/.append style = {draw, rounded corners=0.8},
_BIG_dC_Ir/.append style = {draw, rounded corners=0.8},
_BIG_contract/.append style = {draw, rounded corners=0.8},
_BIG_cont/.append style = {draw, rounded corners=0.8},
_BIG_cert/.append style = {draw, rounded corners=0.8},
_BIG_c/.append style = {draw, rounded corners=0.8}
  ]
    
\node[_BIG_p,  label={[inner sep=0.5, name=n10l]north:{\sf\tiny }}] (n10) {};
\node[_BIG_sCCs, right=1.00 of n10, label={[inner sep=0.5, name=n12l]north:{\sf\tiny SCCs}}] (n12) {};
\node[_BIG_c, right=1.00 of n12, label={[inner sep=0.5, name=n14l]north:{\sf\tiny C($\mathtt{1}$)}}] (n14) {};
\node[_BIG_c, right=1.00 of n14, label={[inner sep=0.5, name=n15l]north:{\sf\tiny C($\mathtt{2}$)}}] (n15) {};
\node[_BIG_c, right=1.00 of n15, label={[inner sep=0.5, name=n16l]north:{\sf\tiny C($\mathtt{3}$)}}] (n16) {};
\node[_BIG_p, right=1.00 of n16, label={[inner sep=0.5, name=n18l]north:{\sf\tiny }}] (n18) {};
\node[_BIG_contract, right=1.00 of n18, label={[inner sep=0.5, name=n19l]north:{\sf\tiny Contract}}] (n19) {};
\node[_BIG_p, right=1.00 of n19, label={[inner sep=0.5, name=n21l]north:{\sf\tiny }}] (n21) {};
\node[_BIG_expiryDate, right=1.00 of n21, label={[inner sep=0.5, name=n23l]north:{\sf\tiny ExpiryDate}}] (n23) {};
\node[_BIG_c, right=1.00 of n23, label={[inner sep=0.5, name=n24l]north:{\sf\tiny C($\mathtt{1}$)}}] (n24) {};
\node[_BIG_c, right=1.00 of n24, label={[inner sep=0.5, name=n25l]north:{\sf\tiny C($\mathtt{2}$)}}] (n25) {};
\node[_BIG_facebookC, below=2.00 of n14, label={[inner sep=0.5, name=n5l]north:{\sf\tiny FB}}] (n5) {};
\node[_BIG_dC_Ir, right=1.00 of n5, label={[inner sep=0.5, name=n6l]north:{\sf\tiny DC}}] (n6) {};
\node[_BIG_dC_Si, right=1.00 of n6, label={[inner sep=0.5, name=n7l]north:{\sf\tiny DC}}] (n7) {};
\node[big site, right=1.0 of n7,](s2){};
\node[_BIG_proc, right=1.50 of s2, label={[inner sep=0.5, name=n1l]north:{\sf\tiny Proc}}] (n1) {};
\node[_BIG_proc, right=1.00 of n1, label={[inner sep=0.5, name=n2l]north:{\sf\tiny Proc}}] (n2) {};
\node[_BIG_cont, right=1.00 of n2, label={[inner sep=0.5, name=n3l]north:{\sf\tiny Cont}}] (n3) {};
\node[big site, right=1.0 of n3,](s1){};
\node[_BIG_cert, fit=(n25)(n25l)(n24)(n24l)(n23)(n23l), label={[inner sep=0.5, name=n22l]north:{\sf\tiny Cert}}] (n22) {};
\node[big site, right=1.0 of n25,](s0){};
\node[_BIG_l, fit=(n19)(n19l)(n18)(n18l), label={[inner sep=0.5, name=n17l]north:{\sf\tiny L($\mathtt{Singapore}$)}}] (n17) {};
\node[_BIG_scheme, fit=(n16)(n16l)(n15)(n15l)(n14)(n14l), label={[inner sep=0.5, name=n13l]north:{\sf\tiny Scheme}}] (n13) {};
\node[_BIG_contract, fit=(n12)(n12l), label={[inner sep=0.5, name=n11l]north:{\sf\tiny Contract}}] (n11) {};

{[on background layer]
\node[_BIG_whatsSystem, fit=(n7)(n7l)(n6)(n6l)(n5)(n5l)(s2), fill=cyan!60!green!13!white, label={[inner sep=0.5, name=n4l]north:{\sf\tiny WhatsSystem}}] (n4) {};}

\node[_BIG_sRType, fit=(n3)(n3l)(n2)(n2l)(n1)(n1l)(s1), label={[inner sep=0.5, name=n0l]north:{\sf\tiny SRType}}] (n0) {};
\node[_BIG_l, fit=(n22)(n22l)(n21)(n21l), label={[inner sep=0.5, name=n20l]north:{\sf\tiny L($\mathtt{China}$)}}] (n20) {};
\node[_BIG_l, fit=(n13)(n13l)(n11)(n11l)(n10)(n10l), label={[inner sep=0.5, name=n9l]north:{\sf\tiny L($\mathtt{Ireland}$)}}] (n9) {};

\node[_BIG_locations, fit=(n20)(n20l)(n17)(n17l)(n9)(n9l) (s0),  label={[inner sep=0.5, name=n8l]north:{\sf\tiny Locations}}] (n8) {};
\node[big region, fit=(n0)(n0l)] (r0) {};
\node[big region, fit=(n4)(n4l)] (r1) {};
\node[big region, fit=(n8)(n8l)] (r2) {};
\coordinate (h0) at ($(n10) + (5.5,-1.0)$);
\draw[big edge, red] (n10) to[out=-90,in=90] (h0);
\draw[big edge,red] (n6) to[out=20,in=-90] (h0);
\draw[big edge,red] (n1) to[out=180,in=-90] (h0);
\coordinate (h1) at ($(n18) + (0.5,-0.3)$);
\draw[big edge, blue] (n18) to[out=0,in=90] (h1);
\draw[big edge, blue] (n7) to[out=0,in=-90] (h1);
\draw[big edge, blue] (n2) to[out=140,in=-90] (h1);
\coordinate (h2) at ($(n21) + (0.3,-0.3)$);
\draw[big edge] (n21) to[out=0,in=90] (h2);
 \draw[big edge] (n5) to[out=50,in=-90] (h2);
\draw[big edge] (n3) to[out=180,in=-90] (h2);
\draw[big edge] (n19) to[out=200,in=-50] (n11);
\draw[big edge ] (n22) to[out=140,in=50] (n13);

\end{tikzpicture}      } 
     \caption{A Partial initial state for the example of transferring data to the data centre in \keyword{Singapore} and to Facebook in \keyword{China}. }   
     \label{fig:2ndIntialState}
\end{figure*}

We introduce an example to show the privacy model's usability and how we can merge it with the specific system. Suppose that the data centre of WhatsApp in \keyword{\mathtt{Ireland}} (\keyword{DC}) needs to transfer users' data to WhatsApp's data centre in \keyword{\mathtt{Singapore}} 
(\keyword{DC}) and to the \keyword{Facebook} branch in \keyword{\mathtt{China}} (\keyword{FB}). ~\cref{fig:2ndIntialState} shows a partial initial state for this example. We consider the \keyword{DC} in \keyword{\mathtt{Singapore}} to be a processor (linked to \keyword{Proc}) that has a \keyword{Contract} with the \keyword{DC} in \keyword{\mathtt{Ireland}}. \keyword{FB} is considered as a data controller (linked to \keyword{Cont}), and the safeguard that it provides is a certification (\keyword{Cert}).

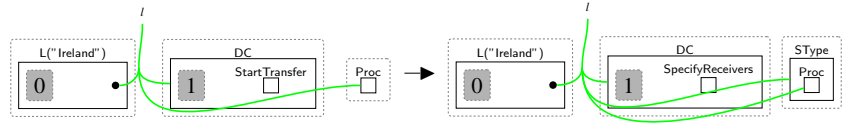
\begin{figure}[]
    \centering
      \resizebox{0.7 \textwidth}{!}{

  \begin{tikzpicture}[
    ,
_BIG_startTransfer/.append style = {draw},
_BIG_specifyReceivers/.append style = {draw},
_BIG_sType/.append style = {draw},
_BIG_proc/.append style = {draw},
_BIG_p/.append style = {fill , circle ,  inner sep=1.3, black},
_BIG_l/.append style = {draw},
_BIG_dC/.append style = {draw}
    ]
    \begin{scope}[local bounding box=lhs, shift={(0,0)}]
      
\node[big site, ] (s2l) {0};
\node[_BIG_p, right=1.00 of s2l, label={[inner sep=0.5, name=n4l]north:{\sf\tiny }}] (n4) {};
\node[big site, right=1.00 of n4,] (s1l) {1};
\node[_BIG_startTransfer, right=1.00 of s1l, label={[inner sep=0.5, name=n2l]north:{\sf\tiny StartTransfer}}] (n2) {};
\node[_BIG_proc, right=1.40 of n2, label={[inner sep=0.5, name=n0l]north:{\sf\tiny Proc}}] (n0) {};
\node[_BIG_l, fit=(n4)(n4l)(s2l), label={[inner sep=0.5, name=n3l]north:{\sf\tiny L("Ireland")}}] (n3) {};
\node[_BIG_dC, fit=(n2)(n2l)(s1l), label={[inner sep=0.5, name=n1l]north:{\sf\tiny DC}}] (n1) {};
\node[big region, fit=(n0)(n0l)] (r0) {};
\node[big region, fit=(n1)(n1l)] (r1) {};
\node[big region, fit=(n3)(n3l)] (r2) {};
\node[] at ($(r0.north west) + (-3.5,0.8)$) (name_l) {\tiny $l$};
\coordinate (h0) at ($(n4) + (0.4,0.3)$);
\draw[big edge] (n4) to[out=0,in=-90] (h0);
\draw[big edge] (n1) to[out=180,in=-90] (h0);
\draw[big edge] (n0) to[out=180,in=-90] (h0);
\draw[big edge] (name_l) to[out=-90,in=90] (h0);

    \end{scope}
    \begin{scope}[local bounding box=rhs, shift={($(lhs.south east) + (1,0)$)}, anchor=south west, scope anchor]
      
\node[big site, ] (s2r) {0};
\node[_BIG_p, right=1.00 of s2r, label={[inner sep=0.5, name=n5l]north:{\sf\tiny }}] (n5) {};
\node[big site, right=1.00 of n5,] (s1r) {1};
\node[_BIG_specifyReceivers, right=1.00 of s1r, label={[inner sep=0.5, name=n3l]north:{\sf\tiny SpecifyReceivers}}] (n3) {};
\node[_BIG_proc, right=1.50 of n3,  label={[inner sep=0.5, name=n1l]north:{\sf\tiny Proc}}] (n1) {};
\node[_BIG_l, fit=(n5)(n5l)(s2r), label={[inner sep=0.5, name=n4l]north:{\sf\tiny L("Ireland")}}] (n4) {};
\node[_BIG_dC, fit=(n3)(n3l)(s1r), label={[inner sep=0.5, name=n2l]north:{\sf\tiny DC}}] (n2) {};
\node[_BIG_sType, fit=(n1)(n1l), label={[inner sep=0.5, name=n0l]north:{\sf\tiny SType}}] (n0) {};
\node[big region, fit=(n0)(n0l)] (r0) {};
\node[big region, fit=(n2)(n2l)] (r1) {};
\node[big region, fit=(n4)(n4l)] (r2) {};
\node[] at ($(r0.north west) + (-3.5,0.3)$) (name_l) {\tiny $l$};
\coordinate (h0) at ($(n5) + (0.5,0.3)$);
\draw[big edge] (n5) to[out=0,in=-90] (h0);
\draw[big edge] (n2) to[out=180,in=-90] (h0);
\draw[big edge] (n1) to[out=200,in=-90] (h0);
\draw[big edge] (n0) to[out=180,in=-90] (h0);
\draw[big edge] (name_l) to[out=-90,in=90] (h0);

    \end{scope}

    \node[xshift=0] at ($(lhs.east)!0.5!(rhs.west)$) {$\rrul$};
  \end{tikzpicture}      } 
     \caption{\rr{dcIrType}: specifying the type of the sender (\keyword{DC} in $\mathtt{Ireland}$).  }   
     \label{rr:dcIrType}
\end{figure}

We need to trigger the privacy model to check the sender and receiver regions and their safeguards. To do so, we should specify the type of the sender (\keyword{SType}) and the receivers (\keyword{RType}) and generate the entity \keyword{CheckReg} via system rules. Rule~\rr{dcIrType} (\cref{rr:dcIrType}) specifies the type of \keyword{DC} in $\mathtt{Ireland}$. As \keyword{DC} is the sender and is linked to \keyword{Proc}, \keyword{SType} is generated around \keyword{Proc} to indicate that the sender is a processor. We also link \keyword{SType} to the sender's pointer in \keyword{L(\mathtt{Ireland})}. The entity \keyword{StartTransfer} is generated by a system rule (shown in \cref{rr:initialisingTrnas} \cref{app:systemRules}). It is nested within \keyword{DC} as \keyword{DC} is the entity initiating the transfer. \keyword{StartTransfer} is replaced with \keyword{SpecifyReceivers} to specify the receivers types before transferring the data.

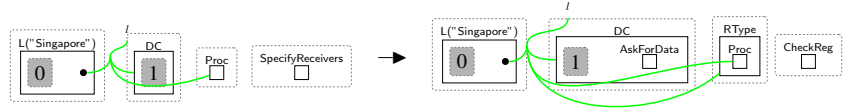
\begin{figure}[]
    \centering
      \resizebox{0.7 \textwidth}{!}{

  \begin{tikzpicture}[
    ,
_BIG_specifyReceivers/.append style = {draw},
_BIG_rType/.append style = {draw},
_BIG_proc/.append style = {draw},
_BIG_p/.append style = {fill , circle ,  inner sep=1.3, black},
_BIG_l/.append style = {draw},
_BIG_dC/.append style = {draw},
_BIG_checkReg/.append style = {draw},
_BIG_askForData/.append style = {draw}
    ]
    \begin{scope}[local bounding box=lhs, shift={(0,0)}]
      
\node[big site, ] (s2l) {0};
\node[_BIG_p, right=0.50 of s2l, label={[inner sep=0.5, name=n4l]north:{\sf\tiny }}] (n4) {};
\node[big site, right=1.00 of n4,] (s1l) {1};

\node[_BIG_dC, fit=(s1l), label={[inner sep=0.5, name=n2l]north:{\sf\tiny DC}}] (n2) {};
\node[_BIG_proc, right=0.80 of s1l , label={[inner sep=0.5, name=n1l]north:{\sf\tiny Proc}}] (n1) {};
\node[_BIG_specifyReceivers, right=1.30 of n1, label={[inner sep=0.5, name=n0l]north:{\sf\tiny SpecifyReceivers}}] (n0) {};
\node[_BIG_l, fit=(n4)(n4l)(s2l), label={[inner sep=0.5, name=n3l]north:{\sf\tiny L("Singapore")}}] (n3) {};
\node[big region, fit=(n0)(n0l)] (r0) {};
\node[big region, fit=(n1)(n1l)] (r1) {};
\node[big region, fit=(n2)(n2l)] (r2) {};
\node[big region, fit=(n3)(n3l)] (r3) {};
\node[] at ($(r0.north west) + (-2.3,0.3)$) (name_l) {\tiny $l$};
\coordinate (h0) at ($(n4) + (0.5,0.3)$);
\draw[big edge] (n4) to[out=0,in=-90] (h0);
\draw[big edge] (n2) to[out=180,in=-90] (h0);
\draw[big edge] (n1) to[out=200,in=-90] (h0);
\draw[big edge] (name_l) to[out=-90,in=90] (h0);

    \end{scope}
    \begin{scope}[local bounding box=rhs, shift={($(lhs.south east) + (1.5,-0.45)$)}, anchor=south west, scope anchor]
      
\node[big site, ] (s2r) {0};
\node[_BIG_p, right=0.50 of s2r, label={[inner sep=0.5, name=n6l]north:{\sf\tiny }}] (n6) {};
\node[big site, right=1.00 of n6,] (s1r) {1};
\node[_BIG_askForData, right=1.00 of s1r, label={[inner sep=0.5, name=n4l]north:{\sf\tiny AskForData}}] (n4) {};

\node[_BIG_proc, right=1.40 of n4 , label={[inner sep=0.5, name=n2l]north:{\sf\tiny Proc}}] (n2) {};
\node[_BIG_checkReg, right=1.0 of n2, label={[inner sep=0.5, name=n0l]north:{\sf\tiny CheckReg}}] (n0) {};
\node[_BIG_l, fit=(n6)(n6l)(s2r), label={[inner sep=0.5, name=n5l]north:{\sf\tiny L("Singapore")}}] (n5) {};
\node[_BIG_dC, fit=(n4)(n4l)(s1r), label={[inner sep=0.5, name=n3l]north:{\sf\tiny DC}}] (n3) {};
\node[_BIG_rType, fit=(n2)(n2l), label={[inner sep=0.5, name=n1l]north:{\sf\tiny RType}}] (n1) {};
\node[big region, fit=(n0)(n0l)] (r0) {};
\node[big region, fit=(n1)(n1l)] (r1) {};
\node[big region, fit=(n3)(n3l)] (r2) {};
\node[big region, fit=(n5)(n5l)] (r3) {};
\node[] at ($(r0.north west) + (-4.0,0.3)$) (name_l) {\tiny $l$};
\coordinate (h0) at ($(n6) + (0.4,0.3)$);
\draw[big edge] (n6) to[out=0,in=-90] (h0);
\draw[big edge] (n3) to[out=180,in=-90] (h0);
\draw[big edge] (n2) to[out=180,in=-90] (h0);
\draw[big edge] (n1) to[out=220,in=-90] (h0);
\draw[big edge] (name_l) to[out=-90,in=90] (h0);

    \end{scope}

    \node[xshift=0] at ($(lhs.east)!0.5!(rhs.west)$) {$\rrul$};
  \end{tikzpicture}      } 
     \caption{\rr{dcSingType}: specifying the type of the receiver (\keyword{DC} in $\mathtt{Singapore}$). }   
     \label{rr:dcSingType}
\end{figure}

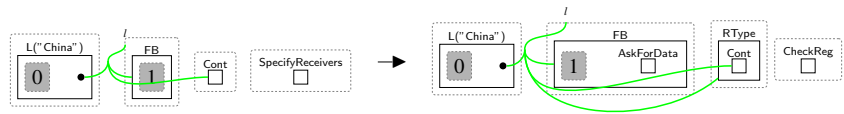
\begin{figure}[]
    \centering
      \resizebox{0.7 \textwidth}{!}{

  \begin{tikzpicture}[
    ,
_BIG_specifyReceivers/.append style = {draw},
_BIG_rType/.append style = {draw},
_BIG_p/.append style = {fill , circle ,  inner sep=1.3, black},
_BIG_l/.append style = {draw},
_BIG_fB/.append style = {draw},
_BIG_cont/.append style = {draw},
_BIG_checkReg/.append style = {draw},
_BIG_askForData/.append style = {draw}
    ]
    \begin{scope}[local bounding box=lhs, shift={(0,0)}]
      
\node[big site, ] (s2l) {0};
\node[_BIG_p, right=0.50 of s2l, label={[inner sep=0.5, name=n4l]north:{\sf\tiny }}] (n4) {};
\node[big site, right=1.00 of n4,] (s1l) {1};
\node[_BIG_fB, fit=(s1l), label={[inner sep=0.5, name=n2l]north:{\sf\tiny FB}}] (n2) {};

\node[_BIG_cont, right=0.80 of s1l, label={[inner sep=0.5, name=n1l]north:{\sf\tiny Cont}}] (n1) {};
\node[_BIG_specifyReceivers, right=1.30 of n1, label={[inner sep=0.5, name=n0l]north:{\sf\tiny SpecifyReceivers}}] (n0) {};
\node[_BIG_l, fit=(n4)(n4l)(s2l), label={[inner sep=0.5, name=n3l]north:{\sf\tiny L("China")}}] (n3) {};
\node[big region, fit=(n0)(n0l)] (r0) {};
\node[big region, fit=(n1)(n1l)] (r1) {};
\node[big region, fit=(n2)(n2l)] (r2) {};
\node[big region, fit=(n3)(n3l)] (r3) {};
\node[] at ($(r0.north west) + (-2.3,0.3)$) (name_l) {\tiny $l$};
\coordinate (h0) at ($(n4) + (0.5,0.3)$);
\draw[big edge] (n4) to[out=0,in=-90] (h0);
\draw[big edge] (n2) to[out=180,in=-90] (h0);
\draw[big edge] (n1) to[out=180,in=-90] (h0);
\draw[big edge] (name_l) to[out=-90,in=90] (h0);

    \end{scope}
    \begin{scope}[local bounding box=rhs, shift={($(lhs.south east) + (1.5,-0.45)$)}, anchor=south west, scope anchor]
      
\node[big site, ] (s2r) {0};
\node[_BIG_p, right=0.50 of s2r, label={[inner sep=0.5, name=n6l]north:{\sf\tiny }}] (n6) {};
\node[big site, right=1.00 of n6,] (s1r) {1};
\node[_BIG_askForData, right=1.00 of s1r, label={[inner sep=0.5, name=n4l]north:{\sf\tiny AskForData}}] (n4) {};

\node[_BIG_cont,right=1.40 of n4, label={[inner sep=0.5, name=n2l]north:{\sf\tiny Cont}}] (n2) {};
\node[_BIG_checkReg, right=1.00 of n2, label={[inner sep=0.5, name=n0l]north:{\sf\tiny CheckReg}}] (n0) {};
\node[_BIG_l, fit=(n6)(n6l)(s2r), label={[inner sep=0.5, name=n5l]north:{\sf\tiny L("China")}}] (n5) {};
\node[_BIG_fB, fit=(n4)(n4l)(s1r), label={[inner sep=0.5, name=n3l]north:{\sf\tiny FB}}] (n3) {};
\node[_BIG_rType, fit=(n2)(n2l), label={[inner sep=0.5, name=n1l]north:{\sf\tiny RType}}] (n1) {};
\node[big region, fit=(n0)(n0l)] (r0) {};
\node[big region, fit=(n1)(n1l)] (r1) {};
\node[big region, fit=(n3)(n3l)] (r2) {};
\node[big region, fit=(n5)(n5l)] (r3) {};
\node[] at ($(r0.north west) + (-4.0,0.3)$) (name_l) {\tiny $l$};
\coordinate (h0) at ($(n6) + (0.4,0.3)$);
\draw[big edge] (n6) to[out=0,in=-90] (h0);
\draw[big edge] (n3) to[out=180,in=-90] (h0);
\draw[big edge] (n2) to[out=180,in=-90] (h0);
\draw[big edge] (n1) to[out=220,in=-90] (h0);
\draw[big edge] (name_l) to[out=-90,in=90] (h0);

    \end{scope}

    \node[xshift=0] at ($(lhs.east)!0.5!(rhs.west)$) {$\rrul$};
  \end{tikzpicture}      } 
     \caption{\rr{FacebookCType}: specifying \keyword{FB} type.   }   
     \label{rr:FacebookCType}
\end{figure}

Rules~\rr{dcSingType} (\cref{rr:dcSingType}) and \rr{FacebookCType} (\cref{rr:FacebookCType}) specify the type of the receivers in \keyword{\mathtt{Singapore}} (\keyword{DC}) and \keyword{\mathtt{China}} (\keyword{FB}), respectively. Since \keyword{DC} is linked to \keyword{Proc}, the \keyword{RType} is generated around \keyword{Proc} and linked to \keyword{DC}'s location. The entity \keyword{AskForData} represents the receiver's request for the data. \keyword{FB} is a data controller, so \keyword{RType} is generated around the token \keyword{Cont} and linked to \keyword{\mathtt{China}}. We also replace \keyword{SpecifyReceivers} with \keyword{CheckReg}.

\begin{figure}[ht]
    \centering
      \resizebox{0.8 \textwidth}{!}{

\begin{tikzpicture}[
  ,
_BIG_sType/.append style = {draw, rounded corners=0.8},
_BIG_rType/.append style = {draw, rounded corners=0.8},
_BIG_ps/.append style = {fill , circle ,  inner sep=1.3, yellow},
_BIG_p'/.append style = {fill , circle ,  inner sep=1.3, blue},
_BIG_l/.append style = {draw, rounded corners=0.8},
_BIG_fB/.append style = {draw, rounded corners=0.8},
_BIG_dC/.append style = {draw, rounded corners=0.8}
  ]
    
\node[big site, ] (s6l) {};
\node[_BIG_p', right=1.00 of s6l, label={[inner sep=0.5, name=n7l]north:{\sf\tiny }}] (n7) {};
\node[big site, right=1.00 of n7,] (s7l) {};
\node[_BIG_p', right=1.00 of s7l, label={[inner sep=0.5, name=n9l]north:{\sf\tiny }}] (n9) {};
\node[big site, right=1.00 of n9,] (s8l) {};
\node[_BIG_ps, right=1.00 of s8l, label={[inner sep=0.5, name=n11l]north:{\sf\tiny }}] (n11) {};
\node[big site, right=1.00 of n11,] (s3l) {};
\node[big site, right=1.00 of s3l,] (s4l) {};
\node[big site, right=1.00 of s4l,] (s5l) {};
\node[big site, right=1.00 of s5l,] (s0l) {};
\node[big site, right=1.00 of s0l,] (s1l) {};
\node[big site, right=1.00 of s1l,] (s2l) {};
\node[_BIG_dC, fit=(s5l), label={[inner sep=0.5, name=n5l]north:{\sf\tiny DC}}] (n5) {};
\node[_BIG_dC, fit=(s4l), label={[inner sep=0.5, name=n4l]north:{\sf\tiny DC}}] (n4) {};
\node[_BIG_fB, fit=(s3l), label={[inner sep=0.5, name=n3l]north:{\sf\tiny FB}}] (n3) {};
\node[_BIG_sType, fit=(s2l), label={[inner sep=0.5, name=n2l]north:{\sf\tiny SType}}] (n2) {};
\node[_BIG_rType, fit=(s1l), label={[inner sep=0.5, name=n1l]north:{\sf\tiny RType}}] (n1) {};
\node[_BIG_rType, fit=(s0l), label={[inner sep=0.5, name=n0l]north:{\sf\tiny RType}}] (n0) {};
\node[_BIG_l, fit=(n11)(n11l)(s8l), label={[inner sep=0.5, name=n10l]north:{\sf\tiny L("Ireland")}}] (n10) {};
\node[_BIG_l, fit=(n9)(n9l)(s7l), label={[inner sep=0.5, name=n8l]north:{\sf\tiny L("Singapore")}}] (n8) {};
\node[_BIG_l, fit=(n7)(n7l)(s6l), label={[inner sep=0.5, name=n6l]north:{\sf\tiny L("China")}}] (n6) {};
\node[big region, fit=(n2)(n2l)(n1)(n1l)(n0)(n0l)] (r0) {};
\node[big region, fit=(n5)(n5l)(n4)(n4l)(n3)(n3l)] (r1) {};
\node[big region, fit=(n10)(n10l)(n8)(n8l)(n6)(n6l)] (r2) {};
\node[] at ($(r0.north west) + (-8.5,0.3)$) (name_l) {\tiny $l$};

\node[right=1.3 of name_l] (name_l') {\tiny $l'$};
\node[right=1.7 of name_l'] (name_l'') {\tiny $l''$};
\coordinate (h0) at ($(n7) + (0.5,0.3)$);
\draw[big edge, red] (n7) to[out=0,in=-90] (h0);
\draw[big edge, red] (n3) to[out=200,in=-90] (h0);
\draw[big edge, red] (n0) to[out=180,in=-90] (h0);
\draw[big edge, red] (name_l) to[out=-90,in=90] (h0);
\coordinate (h1) at ($(n9) + (0.5,0.3)$);
\draw[big edge, black] (n9) to[out=0,in=-90] (h1);
\draw[big edge, black] (n4) to[out=200,in=-90] (h1);
\draw[big edge, black] (n1) to[out=180,in=-90] (h1);
\draw[big edge, black] (name_l') to[out=-90,in=90] (h1);
\coordinate (h2) at ($(n11) + (0.6,0.3)$);
\draw[big edge] (n11) to[out=0,in=-90] (h2);
\draw[big edge] (n5) to[out=180,in=-90] (h2);
\draw[big edge] (n2) to[out=180,in=-90] (h2);
\draw[big edge] (name_l'') to[out=-90,in=90] (h2);

\end{tikzpicture}      } 
     \caption{ The tagged pointers after specifying \keyword{DC} in \keyword{\mathtt{Ireland}} and \keyword{\mathtt{Singapore}} and \keyword{FB} locations.  }   
     \label{fig:taggedPointers}
\end{figure}
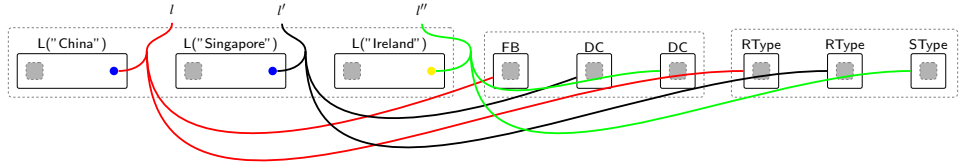

After specifying the sender and receivers types and generating \keyword{CheckReg}, we can start checking the regions by tagging the pointers linked to the specified types. We specify the region of the sender \keyword{DC} by using rule \rr{checkSType} (\cref{rr:checkSType}).
We also use rule \rr{checkRType} (\cref{rr:checkRType}) to check the region of the receivers \keyword{DC} and \keyword{FB}.
By applying these rules, the pointers that will be tagged are the pointers that are located in $\keyword{L(\mathtt{China)}}$, $\keyword{L(\mathtt{Singapore})}$ and $\keyword{L(\mathtt{Ireland})}$ as shown in \cref{fig:taggedPointers}. We can observe that the tagged pointers are located in different countries, so the transfer is restricted.

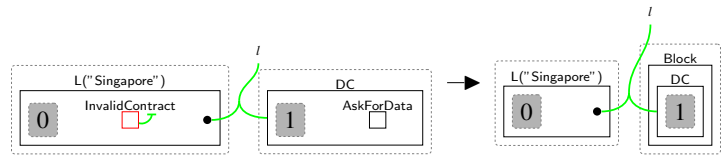
\begin{figure}[]
    \centering
      \resizebox{0.6 \textwidth}{!}{

  \begin{tikzpicture}[
    ,
_BIG_p/.append style = {fill , circle ,  inner sep=1.3, black},
_BIG_l/.append style = {draw},
_BIG_invalidContract/.append style = {draw, red},
_BIG_dC/.append style = {draw},
_BIG_block/.append style = {draw},
_BIG_askForData/.append style = {draw}
    ]
    \begin{scope}[local bounding box=lhs, shift={(0,0)}]
      
\node[big site, ] (s1l) {0};
\node[_BIG_invalidContract, right=1.00 of s1l, label={[inner sep=0.5, name=n3l]north:{\sf\tiny InvalidContract}}] (n3) {};
\node[_BIG_p, right=1.00 of n3, label={[inner sep=0.5, name=n4l]north:{\sf\tiny }}] (n4) {};
\node[big site, right=1.00 of n4,] (s0l) {1};
\node[_BIG_askForData, right=1.00 of s0l, label={[inner sep=0.5, name=n1l]north:{\sf\tiny AskForData}}] (n1) {};
\node[_BIG_l, fit=(n4)(n4l)(n3)(n3l)(s1l), label={[inner sep=0.5, name=n2l]north:{\sf\tiny L("Singapore")}}] (n2) {};
\node[_BIG_dC, fit=(n1)(n1l)(s0l), label={[inner sep=0.5, name=n0l]north:{\sf\tiny DC}}] (n0) {};
\node[big region, fit=(n0)(n0l)] (r0) {};
\node[big region, fit=(n2)(n2l)] (r1) {};
\node[] at ($(r0.north west) + (0,0.3)$) (name_l) {\tiny $l$};
\draw[big edgec] (n3) to[out=-20,in=-90] ($(n3.0) + (0.2,0.1)$);
\coordinate (h0) at ($(n4) + (0.5,0.3)$);
\draw[big edge] (n4) to[out=0,in=-90] (h0);
\draw[big edge] (n0) to[out=180,in=-90] (h0);
\draw[big edge] (name_l) to[out=-90,in=90] (h0);

    \end{scope}
    \begin{scope}[local bounding box=rhs, shift={($(lhs.south east) + (1,0)$)}, anchor=south west, scope anchor]
      
\node[big site, ] (s1r) {0};
\node[_BIG_p, right=0.80 of s1r, label={[inner sep=0.5, name=n3l]north:{\sf\tiny }}] (n3) {};
\node[big site, right=1.00 of n3,] (s0r) {1};
\node[_BIG_dC, fit=(s0r), label={[inner sep=0.5, name=n1l]north:{\sf\tiny DC}}] (n1) {};
\node[_BIG_l, fit=(n3)(n3l)(s1r), label={[inner sep=0.5, name=n2l]north:{\sf\tiny L("Singapore")}}] (n2) {};
\node[_BIG_block, fit=(n1)(n1l), label={[inner sep=0.5, name=n0l]north:{\sf\tiny Block}}] (n0) {};
\node[big region, fit=(n0)(n0l)] (r0) {};
\node[big region, fit=(n2)(n2l)] (r1) {};
\node[] at ($(r0.north west) + (0,0.3)$) (name_l) {\tiny $l$};
\coordinate (h0) at ($(n3) + (0.5,0.3)$);
\draw[big edge] (n3) to[out=0,in=-90] (h0);
\draw[big edge] (n1) to[out=180,in=-90] (h0);
\draw[big edge] (name_l) to[out=-90,in=90] (h0);

    \end{scope}

    \node[xshift=0] at ($(lhs.east)!0.5!(rhs.west)$) {$\rrul$};
  \end{tikzpicture}      } 
     \caption{\rr{preventSing}: preventing the transfer to the \keyword{DC} in \keyword{\mathtt{Singapore}}.  }   
     \label{rr:preventSing}
\end{figure}

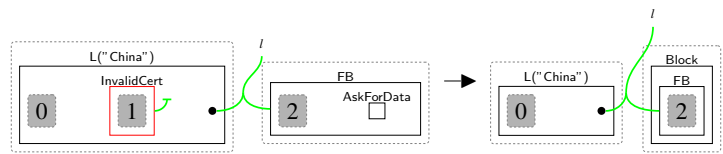
\begin{figure}[]
    \centering
      \resizebox{0.6 \textwidth}{!}{

  \begin{tikzpicture}[
    ,
_BIG_p/.append style = {fill , circle ,  inner sep=1.3, black},
_BIG_l/.append style = {draw},
_BIG_invalidCert/.append style = {draw,red},
_BIG_fB/.append style = {draw},
_BIG_block/.append style = {draw},
_BIG_askForData/.append style = {draw}
    ]
    \begin{scope}[local bounding box=lhs, shift={(0,0)}]
      
\node[big site, ] (s1l) {0};
\node[big site, right=1.00 of s1l,] (s2l) {1};
\node[_BIG_p, right=1.00 of s2l, label={[inner sep=0.5, name=n4l]north:{\sf\tiny }}] (n4) {};
\node[big site, right=1.00 of n4,] (s0l) {2};
\node[_BIG_askForData, right=1.00 of s0l, label={[inner sep=0.5, name=n1l]north:{\sf\tiny AskForData}}] (n1) {};
\node[_BIG_invalidCert, fit=(s2l), label={[inner sep=0.5, name=n3l]north:{\sf\tiny InvalidCert}}] (n3) {};
\node[_BIG_l, fit=(n4)(n4l)(n3)(n3l)(s1l), label={[inner sep=0.5, name=n2l]north:{\sf\tiny L("China")}}] (n2) {};
\node[_BIG_fB, fit=(n1)(n1l)(s0l), label={[inner sep=0.5, name=n0l]north:{\sf\tiny FB}}] (n0) {};
\node[big region, fit=(n0)(n0l)] (r0) {};
\node[big region, fit=(n2)(n2l)] (r1) {};
\node[] at ($(r0.north west) + (0,0.3)$) (name_l) {\tiny $l$};
\draw[big edgec] (n3) to[out=0,in=-90] ($(n3.0) + (0.2,0.2)$);
\coordinate (h0) at ($(n4) + (0.5,0.3)$);
\draw[big edge] (n4) to[out=0,in=-90] (h0);
\draw[big edge] (n0) to[out=180,in=-90] (h0);
\draw[big edge] (name_l) to[out=-90,in=90] (h0);

    \end{scope}
    \begin{scope}[local bounding box=rhs, shift={($(lhs.south east) + (1,0)$)}, anchor=south west, scope anchor]
      
\node[big site, ] (s1r) {0};
\node[_BIG_p, right=1.00 of s1r, label={[inner sep=0.5, name=n3l]north:{\sf\tiny }}] (n3) {};
\node[big site, right=1.00 of n3,] (s0r) {2};
\node[_BIG_fB, fit=(s0r), label={[inner sep=0.5, name=n1l]north:{\sf\tiny FB}}] (n1) {};
\node[_BIG_l, fit=(n3)(n3l)(s1r), label={[inner sep=0.5, name=n2l]north:{\sf\tiny L("China")}}] (n2) {};
\node[_BIG_block, fit=(n1)(n1l), label={[inner sep=0.5, name=n0l]north:{\sf\tiny Block}}] (n0) {};

\node[big region, fit=(n0)(n0l)] (r0) {};
\node[big region, fit=(n2)(n2l)] (r1) {};
\node[] at ($(r0.north west) + (0,0.3)$) (name_l) {\tiny $l$};
\coordinate (h0) at ($(n3) + (0.4,0.3)$);
\draw[big edge] (n3) to[out=0,in=-90] (h0);
\draw[big edge] (n1) to[out=180,in=-90] (h0);
\draw[big edge] (name_l) to[out=-90,in=90] (h0);

    \end{scope}

    \node[xshift=0] at ($(lhs.east)!0.5!(rhs.west)$) {$\rrul$};
  \end{tikzpicture}      } 
     \caption{\rr{preventChina}: preventing the transfer to the \keyword{FB} in \keyword{\mathtt{China}}.   }   
     \label{rr:preventChina}
\end{figure}

Based on \cref{fig:2ndIntialState}, \keyword{Singapore} and \keyword{China} are not specified as adequate countries, \ie no pointers in \keyword{Singapore} or \keyword{China} linked to \keyword{Adeq}, as is the case with the \keyword{\mathtt{US}} in \cref{WhatsApp_initialState}, meaning that the safeguards must be checked. The entity \keyword{Contract} that is nested within $\keyword{L(\mathtt{Singapore})}$ does not include \keyword{SCCs}. This indicates that \keyword{DC} in  $\keyword{\mathtt{Singapore}}$ does not incorporate the \keyword{SCCs} in their contract, \eg does not agree to adhere to the \keyword{SCCs}.
Rule \rr{checkingSCCs} (\cref{rr:checkingSCCs}) is applied to tag the \keyword{Contract} as \keyword{InvalidContract} and close its link connected to the entity \keyword{Contract} in $\keyword{\mathtt{Ireland}}$. The entity \keyword{InvalidContract} is used as a flag to prevent the transfer and block the \keyword{DC} in $\keyword{\mathtt{Singapore}}$ through the system rule \rr{preventSing}~(\cref{rr:preventSing}). Similarly, \keyword{FB} uses the certification mechanism that has only two criteria ($\keyword{C\mathtt{(1)}}$ and $\keyword{C\mathtt{(2)}}$) as shown in ~\cref{fig:2ndIntialState}. By applying rule \rr{tagCriteria(x)} (\cref{rr:tagCriteria}), only two criteria are tagged. In this case, rule \rr{InvalidCert}~(\cref{rr:tagInvalidCert}) tags the \keyword{Cert} as \keyword{InvalidCert}. To block \keyword{FB}, we define a system rule that matches explicitly on \keyword{InvalidCert} as shown in~\cref{rr:preventChina}. Because we prevented the transfer to \keyword{DC} and \keyword{FB}, we can safely discard the entities \keyword{InvalidContract} and \keyword{InvalidCert}.   

\section{Verification}
\label{sec:verification}

The aim of defining the privacy model is to generate a transition system used to conduct the \textit{verification} step, \ie proving the GDPR requirements for international data transfer within the system. BigraphER utilises the initial state, \eg \cref{WhatsApp_initialState}, and the reaction rules (privacy rules and system rules) to \textit{automatically} generate the transition system. The produced transition system consists of states (bigraphs) that describe the system configurations. Each state is a result of applying a rule (transition).
Manually checking privacy properties, \ie the GDPR requirements, within the transition system is challenging as the number of states can be considerable. For example, in the WhatsApp example, the number of the generated states is 193, while the number of the transitions is 267. This complexity would increase if we model a more complex system with additional reaction rules or locations.

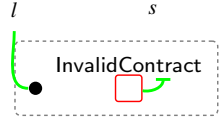
\begin{figure}[]
    \centering
      \resizebox{0.2 \textwidth}{!}{
       \begin{tikzpicture}[
  ,
_BIG_p'/.append style = {fill , circle ,  inner sep=1.3, black},
_BIG_invalidContract/.append style = {draw, red, rounded corners=0.8}
  ]
    
\node[_BIG_p',  label={[inner sep=0.5, name=n0l]north:{\sf\tiny }}] (n0) {};
\node[_BIG_invalidContract, right=0.7 of n0, label={[inner sep=0.5, name=n1l]north:{\sf\tiny InvalidContract}}] (n1) {};
\node[big region, fit=(n1)(n1l)(n0)(n0l)] (r1) {};
\node[] at ($(r1.north west) + (0,0.3)$) (name_a) {\tiny $l$};
\node[right=1 of name_a] (name_s) {\tiny $s$};
\draw[big edgec] (n1) to[out=-10,in=-90] ($(n1.0) + (0.2,0.1)$);
\draw[big edge] (n0) to[out=180,in=-90] (name_a);

\end{tikzpicture}      } 
     \caption{Bigraph pattern of the \pred{invalidContra} predicate.}   
     \label{pred:invalidContra}
\end{figure}

Model checkers are tools used to \textit{automatically} prove properties within the transition system. We use the PRISM~\cite{KNP11} model checker as BigraphER supports it. To use PRISM, the properties that we aim to check should be encoded using a logic language. We use a non-probabilistic temporal logic language as the requirements we aim to check are specified based on \textit{regulations}. In particular, we use Computation Tree Logic (CTL)~\cite{clarke1981design} to encode the GDPR requirements for international data transfer because CTL quantifies the properties over all paths. This allows us to prove that the GDPR requirements hold across all possible execution paths. We also define a set of predicates using BigraphER to label the states. These predicates represent the right-hand sides of the reaction rules, and PRISM utilises them to parse the transition system. For instance, we define predicate \pred{invalidContra} (\cref{pred:invalidContra}) to label the states that have the match of the right-hand side of rule \rr{checkingSCCs}~(\cref{rr:checkingSCCs}).

The first requirement that we need to check is the adequacy requirement: \textit{no data transfer if the country is inadequate}. The CTL syntax of this requirement is:
\begin{equation}
\mathbf{A} \left[\, \mathbf{G} \ (\ \neg \pred{adequateCountry} \implies  \ \neg \pred{dataTransfer})  \, \right]
\end{equation}
\pred{adequateCountry} is a predicate that labels all states that contain the match of the right-hand side of rule \rr{checkingAdeq} (\cref{rr:checkingAdeq}). The second predicate is \pred{dataTransfer}, which labels the states that transfer the data to the receiver via system rules. $\mathbf{A}$ is a path quantifier, meaning \textit{for all paths}. $\mathbf{G}$ is a state quantifier and it means \textit{for all states} (globally). $\mathbf{AG}$ means the property should hold for all paths globally throughout the entire transition system. Here, we check that: for all paths ($\mathbf{A}$), if \pred{adequateCountry} does not hold ($\neg$), \pred{dataTransfer} should not hold ($\neg$) in all states along that path ($\mathbf{G}$).

The second requirement that we need to check is: \textit{ no data transfer if the SCCs are invalid}:
\begin{equation}
\mathbf{A} \left[\, \mathbf{G}\, ( \pred{invalidContra} \implies \neg \pred{dataTransfer} ) \, \right]
\end{equation}
As previously mentioned, \pred{invalidContra} is a predicate that holds if the \keyword{Contract} is invalid, so in this case we check: for all paths of the transition system, if \pred{invalidContra} holds in a state, \pred{dataTransfer} should not hold in the same state throughout the path.

The third requirement is: \textit{no transfer should be performed if the certification is invalid}:   
\begin{equation}
\mathbf{A} \left[\, \mathbf{G}\, ( \pred{invalidCert} \implies \neg \pred{dataTransfer} ) \, \right]
\end{equation}
This property proves that: it is always the case that if the certification is invalid, \eg expired or does not meet the criteria, the transfer is never performed.

The last requirement is: \textit{no transfer should be performed if the certification is withdrawn}: 
\begin{equation}
\mathbf{A} \left[\, \mathbf{G}\, ( \pred{withdrawCert} \implies \ ( \mathbf{X} \ \neg \ \pred{dataTransfer}) ) \, \right]
\end{equation}
Here, we prove that for all paths, if \pred{withdrawCert} is true, then \pred{dataTransfer} should be false in all subsequent states on that path. Since \pred{withdrawCert} only becomes true when a system rule generates \keyword{WithdReq}, as discussed in \cref{subSec:WithdCert}, we apply the next-state operator $\mathbf{X}$ to ensure that \pred{dataTransfer} is false starting from the state immediately following the state that holds \pred{withdrawCert}. This use of $\mathbf{X}$ reflects the system’s sequential processing order.

\section{Sorting}
\label{sec:sorts}
Sorting schemes ensure that bigraphs are well-formed by imposing constraints that prevent undesired or nonsensical models \cite{milner2009space}, \eg preventing scenarios where $\keyword{Adeq}$ is incorrectly nested within \keyword{Scheme}, or where \keyword{Cert} is linked to \keyword{FB} instead of \keyword{Scheme}\footnote{Currently, no automated tool supports these sorting schemes.}.

We use the notation and grammar presented in \cite{archibald.sevegnangi_BigraphsPaperOfSorts} to define our sorts. Sorts can take two forms: $\keyword{\mathtt{sort \ y}}$ or $\keyword{\mathtt{sort \ b = A}}$. The former specifies the sort of link ports, while the latter indicates that \keyword{A} is an atomic entity with sort $\keyword{\mathtt{b}}$.
If the entity \keyword{A} is linked to other entities and has children, its sort pattern is defined as: 
\begin{equation*}
\keyword{\mathtt{sort \ b \ = A \{y \rightarrow z^*\} \ x \ + \ w}}
\end{equation*}
The sort $\keyword{\mathtt{y}}$ represents the sort of the port on \keyword{A} that connects to zero or more ports of sort $\keyword{\mathtt{z}}$. Meanwhile, $\keyword{\mathtt{x}}$ and $\keyword{\mathtt{w}}$ refer to the sorts of \keyword{A}'s children, \ie \keyword{A} has children of sort $\keyword{\mathtt{x}}$ \textit{or} $\keyword{\mathtt{w}}$. \cref{tab:Sorts_notations} describes through some examples the sort notations used in this paper. 
 
\begin{table} 
    \centering
    \caption{Description of sort notations.}
    \begin{tabularx}{\textwidth}{>{\centering\arraybackslash}p{0.15\textwidth} >{\centering\arraybackslash}p{0.25\textwidth} X}
       
         \textbf{Notation} & \textbf{Example} & \textbf{Description} \\ 
         \midrule
     $s^*$      & $A \ s^*$        & The entity $A$ has \textit{zero} or more children of sort $s$. \\
    & \vspace{10pt} & \\
     $s\times s^*$ & $A \ s \times s^*$ & The entity $A$ has at least \textit{one} child of sort $s$. \\
     & \vspace{10pt} & \\
     $s\times i$ & $A\{a \rightarrow s \times i$\} & The entity $A$ has a port of sort $a$, which is connected to ports of sorts $s$ AND $i$. \\
     & \vspace{10pt} & \\
     $s+i$ & $A\{a \rightarrow s + i$\} & The entity $A$ has a port of sort $a$, which is connected to a port of sort $s$ OR $i$. \\
     & \vspace{10pt} & \\
     $s+1$ & $A \ s + 1$ & The entity $A$ has children of sort $s$ OR it does not have any (\ie it is empty).\\
    \end{tabularx}
    \label{tab:Sorts_notations}
\end{table}

To accurately model privacy and system entities, along with their associated behaviours, we define the necessary sorts for maintaining the integrity of the model and preventing ill-structured configurations across all perspectives.

\paragraph{SRType Perspective.} We define the following sorts to ensure a well-formed perspective:
\begin{center}
\begin{tabular}{l} 
$\keyword{\mathtt{sort \ a, \ sort \ p, \ sort \ et, \ sort\  tp, \ sort \ sy }}$ \\
\keyword{\mathtt{sort \ sr \ = \  Cont\{a \rightarrow  (p + tp) \times \ sy + et \}  }} \\
\ \ \ \ \ \ \ \ \ \ \ \ \ \ \ \ \ \ \ \ 
$\mid$ \ \keyword{\mathtt{ Proc\{a \rightarrow  (p + tp)  \ \times \ sy + et \}  }} 
   \\
 $\keyword{\mathtt{sort \ ent \ = SType \{et \rightarrow (p+tp) \times a \times sy\} \ sr}} \  \mid \ \keyword{\mathtt{ RType \{et \rightarrow (p+tp)\times a \times sy\} \ sr}}
 $\\  
\keyword{\mathtt{ sort \ srt \ = SRType \ sr\ \times \ sr^* }} \\ 
\end{tabular}
\end{center}

The entities \keyword{Cont} and \keyword{Proc}, both having the sort \keyword{\mathtt{sr}}, each have a port of sort \keyword{\mathtt{a}}. This port is linked to:
\begin{itemize}
    \item Either a port of sort \keyword{\mathtt{p}} (for an untagged pointer) or \keyword{\mathtt{tp}} (for a tagged pointer) along with a system entity via a port of sort \keyword{\mathtt{sy}},
    \item Or to a port of sort \keyword{\mathtt{et}} on the entity \keyword{SType} or \keyword{RType}.
\end{itemize}

The sort \keyword{\mathtt{ent}} represents the sort of \keyword{SType} and \keyword{RType}. As these entities are used to specify the type of the sender and receiver (whether \keyword{Cont} or \keyword{Proc}), each must include at least one child of sort \keyword{\mathtt{sr}}. Both \keyword{SType} and \keyword{RType} are linked to three ports: one on a tagged or untagged pointer, one on the entity type, and one on the system entity.
The entity \keyword{SRType}, with sort \keyword{\mathtt{srt}}, acts as the parent of \keyword{Cont} and \keyword{Proc}. This imposes a structural constraint where \keyword{SRType} contains one or more entities of sort \keyword{\mathtt{sr}}, as illustrated in Figures \ref{WhatsApp_initialState} and \ref{fig:2ndIntialState}.

By defining and applying these sorts, we ensure that site 0, nested within \keyword{SType} and \keyword{RType} in rule \rr{checkSType} and \rr{checkRType} (\cref{rr:checkingReg_All}), either represents the entity \keyword{Cont} or \keyword{Proc}. We also ensure that  \keyword{SType} and \keyword{RType} are linked to a pointer, with the other end of the link connected to a port on the system entity.

\paragraph{Locations Perspective.}
Similar to the \keyword{SRType} perspective, we define the following sorts for the entities of the \keyword{Locations} perspectives:
\begin{center}
\begin{tabular}{l} 
$\keyword{\mathtt{sort \ c, \ sort \ i, \ sort \ t, \ sort\ p, \ sort \ st, sort \ d}}$ \\
$\keyword{\mathtt{sort \ cr \ = \ C(1) \mid C(2) \mid C(3) }}$ \\
$\keyword{\mathtt{sort \ tcr \ = \ C'(1) \mid C'(2) \mid C'(3)}}$ \\
$\keyword{\mathtt{sort \ e \ = \ ExpiryDate}}$ \\
$\keyword{\mathtt{sort \  g \ = \ Greater \  e }}$ 
\\
$\keyword{\mathtt{sort \ s \ = \ SCCs}}$ \\ 
$\keyword{\mathtt{sort \ ad \ = Adeq\{d \rightarrow p \times p^*\} }}$ \\
$\keyword{\mathtt{sort \ pnt \ = \ P\{p \rightarrow (sy \times a) + d +et \} \ \mid \ P'\{tp \rightarrow (sy \times a) + et \}  \ \mid \ Ps\{tp \rightarrow (sy \times a) + et \}  }}$
\\ 
$\keyword{\mathtt{sort \ scm \ = Scheme \{ sc \rightarrow c \times c^* \} \ cr \times cr^* }}$
\\ 
$\keyword{\mathtt{sort \ certf \ = Cert \{ c \rightarrow sc\}  \ (cr^* + tcr^*) \ \times e }}$\\
           $ \ \ \ \ \ \ \ \ \ \ \ \ \ \ \ \ \ \ \ \ \ \ \ \ \ 
            \mid \keyword{\mathtt{InvalidCert\{ c \rightarrow 1\}  \ tcr^* \ \times e }}$ \\ 
          $ \ \ \ \ \ \ \ \ \ \ \ \ \ \ \ \ \ \ \ \ \ \ \ \ \ \mid \keyword{\mathtt{CompliantCert\{ c \rightarrow sc\} \ ( tcr \times  tcr^* \ ) \ \times e }} $ 
\\               
$ \keyword{\mathtt{ sort\ ctr \ = \ Contract \{ t \rightarrow t \times t^* \} \ s +1  }} $ 
\\
$ \keyword{\mathtt{ sort\ inctr \ =  InvalidContract \{ i \rightarrow 1 \}}}$
\\ 
$\keyword{\mathtt{sort \ l \ = L \ (Ireland)\ (pnt^* \times pnt) \times \ (ad \ + \ sc + \ ctr) }} $ \\
       $\ \ \ \ \ \ \ \ \ \ \ \ \ \ \ \ \ \mid \keyword{\mathtt{ L \ (US)\ (pnt^* \times pnt)}} $\\ 
        $ \ \ \ \ \ \ \ \ \ \ \ \ \ \ \ \ \  \mid \keyword{\mathtt{ L \ (Mexico)\ (pnt^* \times pnt) \ \times \ (ctr \ + \  inctr) }} $\\
        $ \ \ \ \ \ \ \ \ \ \ \ \ \ \ \ \ \  \mid \keyword{\mathtt{ L \ (Singapore)\ (pnt^* \times pnt) \ \times \ (ctr \ + \  inctr) }} $\\
        $ \ \ \ \ \ \ \ \ \ \ \ \ \ \ \ \ \ \mid \keyword{\mathtt{ L \ (Dubai)\ (pnt^* \times pnt) \ \times \ certf }} $ \\
        $ \ \ \ \ \ \ \ \ \ \ \ \ \ \ \ \ \ \mid \keyword{\mathtt{ L \ (China)\ (pnt^* \times  pnt) \ \times \ certf }} $
\\
$\keyword{\mathtt{sort \ loc \ = Locations\ l^* \times l }}$ 
\end{tabular}
\end{center}

We assign the sort \keyword{\mathtt{pnt}} to the pointers \keyword{P}, \keyword{P'}, and \keyword{Ps}, which represent the solid black, blue, and yellow bullets, respectively. Pointer \keyword{P} has a port of sort \keyword{\mathtt{p}}, while pointers \keyword{P'} and \keyword{Ps} have ports of sort \keyword{\mathtt{tp}}. These pointers are linked to two ports: one on a system entity (with sort \keyword{\mathtt{sy}}) and one on a corresponding entity type within the \keyword{SRType} perspective via a port of sort \keyword{\mathtt{a}}.
Additionally, pointer \keyword{P} can be linked to a port of sort \keyword{\mathtt{d}} on the entity \keyword{Adeq}, or to \keyword{SType} or \keyword{RType} via a port of sort \keyword{\mathtt{et}}, while pointers \keyword{P'} and \keyword{Ps} should not be linked to a port of sort \keyword{\mathtt{d}}.

The entity \keyword{Scheme} is linked to one or more certifications (\keyword{Cert}) and must always include at least one criterion of sort \keyword{\mathtt{cr}} or \keyword{\mathtt{tcr}}. The certifications, whether invalid or compliant, are assigned the sort \keyword{\mathtt{certf}}. Both \keyword{Cert} and \keyword{CompliantCert} should be linked to a port on the entity \keyword{Scheme}, and both have a child of sort \keyword{\mathtt{e}}. However, \keyword{Cert} may contain zero or more \textit{untagged} or \textit{tagged} criteria, while \keyword{CompliantCert} must always contain at least one \textit{tagged} criterion to satisfy the criteria specified by the \keyword{Scheme}. In contrast, \keyword{InvalidCert} can include zero or more tagged criteria. As shown in \cref{rr:tagInvalidCert} and \cref{rr:ExpiredDate}, \keyword{InvalidCert} has a closed link, denoted by $1$ in the defined sorts, indicating the absence of a connection to \keyword{Scheme}.

The sort of \keyword{Contract} is \keyword{\mathtt{ctr}}. Since \keyword{Contract} is linked to one or more contracts, the output and input ports share the same sort (\keyword{\mathtt{t}}). \keyword{Contract} may either have a child of sort \keyword{\mathtt{s}}, or it may be empty, in which case it can be tagged as \keyword{InvalidContract} by rule \rr{checkingSCCs} (\cref{rr:checkingSCCs}), which also closes its link.

The parameterised entity \keyword{L} is defined with the sort \keyword{\mathtt{l}}. The sorts of its children vary based on its parameters. For example, \keyword{L\mathtt{(Ireland)}} is the sender country and must contain a child of sort \keyword{\mathtt{ad}}, \keyword{\mathtt{sc}}, or \keyword{\mathtt{ctr}} to allow the GDPR compliance check.

Receiver countries that use \keyword{SSCs}, \eg \keyword{L\mathtt{(Mexico)}} and \keyword{L\mathtt{(Singapore)}}, must include the entity \keyword{Contract} (sort \keyword{\mathtt{ctr}}) or \keyword{InvalidContract} (sort \keyword{\mathtt{inctr}}) if the contract is empty, \ie it does not contain \keyword{SCCs}. In contrast, countries like \keyword{L\mathtt{(Dubai)}} and \keyword{L\mathtt{(China)}} use a certification mechanism, so their children must be of sort \keyword{\mathtt{certf}}. If the receiver country is adequate, it does not contain any safeguards, meaning it includes only tagged or untagged pointers.

Regardless of the value of the parameter for entity \keyword{L}, it must always contain at least one pointer (tagged or untagged) that links to system entities and their types. The parent of \keyword{L} is the entity \keyword{Locations} that has sort \keyword{\mathtt{loc}} and it should contain at least one child of sort \keyword{\mathtt{l}}.

\paragraph{WhatsSystem Perspective.} The system consists of various entities such as \keyword{DC}, \keyword{FB}, and others, each of which has specific roles and children entities based on their functions in the data exchange process:
\begin{center}
\begin{tabular}{l} 
$\keyword{\mathtt{sort \ snd \ = SpecifyReceivers \ \mid CheckReg \mid \ SameRegion \ \mid \ Adequate}}$ \\
$\keyword{\mathtt{sort \ req \ = \ AskForData \ \mid \  WithdReq }}$ \\
$\keyword{\mathtt{sort \ info \ = CopyInfo}}$ \\ 
$\keyword{\mathtt{sort \ cd \ = \ CurrentDate}}$  \\
$\keyword{\mathtt{sort \ chex \ = \ CheckExp\ cd}}$  \\
\keyword{\mathtt{sort \ sys \ = \  DC\{sy \rightarrow (p+tp) \times \ a + et\} \ snd + req + info + chex +  1}} \\
\ \ \ \ \ \ \ \ \ \ \ \ \ \ \ \ \ \ \ \ \ \  $\mid$ \keyword{\mathtt{FB\{sy \rightarrow (p+tp) \times \ a + et\}\ req + info + 1}}\\
 $\keyword{\mathtt{sort \ b \ = \ Block \ sys}}$ \\ 
\end{tabular}
\end{center}

\keyword{DC} and \keyword{FB} are connected by a hyper-edge to a tagged or untagged pointer, as well as to their corresponding types, \eg \keyword{Cont}. When the entities \keyword{SType} and \keyword{RType} are generated, they can be linked to \keyword{DC} or \keyword{FB} via a port of sort \keyword{\mathtt{et}}.
In our model, \keyword{DC} can act as both a sender and a receiver. When it is a sender, it contains the following entities: \keyword{SpecifyReceivers}, \keyword{CheckReg} or \keyword{SameRegion}. When \keyword{DC} is a receiver, it can contain \keyword{AskForData} or \keyword{WithdReq}. 
\keyword{FB} and Other companies in the model, \eg \keyword{Whats}, \keyword{Meta} \etc, contains children such as \keyword{AskForData} or \keyword{CopyInfo}, which are specifically related to their role in requesting and receiving data. 
The entity \keyword{Block} is used to block companies from accessing the data, thereby always containing one child of sort \keyword{\mathtt{sys}}.

By applying these sorts to any initial state modelling the starting configuration of a system and all the reaction rules, we can formally verify the same properties proven by PRISM as discussed in \cref{sec:verification}. For example, we ensure that data transfer occurs only if the \keyword{Contract} is valid, specifically when it includes \keyword{SCCs}. Without the use of these sorts, we risk the \keyword{Contract} containing unintended entities instead of \keyword{SCCs}, which would mean it fails to be marked as \keyword{InvalidContract} and could wrongly allow the transfer to proceed.

\section{Privacy Violations Detection}
\label{sec:violation_ex}
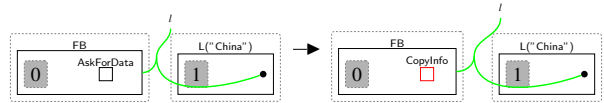
\begin{figure}[]
    \centering
      \resizebox{0.5 \textwidth}{!}{

  \begin{tikzpicture}[
    ,
_BIG_p/.append style = {fill , circle ,  inner sep=1.3, black},
_BIG_l/.append style = {draw},
_BIG_fB/.append style = {draw},
_BIG_copyInfo/.append style = {draw,red},
_BIG_askForData/.append style = {draw}
    ]
    \begin{scope}[local bounding box=lhs, shift={(0,0)}]
      
\node[big site, ] (s1l) {0};
\node[_BIG_askForData, right=1.00 of s1l, label={[inner sep=0.5, name=n3l]north:{\sf\tiny AskForData}}] (n3) {};
\node[big site, right=1.40 of n3,] (s0l) {1};
\node[_BIG_p, right=1.00 of s0l, label={[inner sep=0.5, name=n1l]north:{\sf\tiny }}] (n1) {};
\node[_BIG_fB, fit=(n3)(n3l)(s1l), label={[inner sep=0.5, name=n2l]north:{\sf\tiny FB}}] (n2) {};
\node[_BIG_l, fit=(n1)(n1l)(s0l), label={[inner sep=0.5, name=n0l]north:{\sf\tiny L("China")}}] (n0) {};
\node[big region, fit=(n0)(n0l)] (r0) {};
\node[big region, fit=(n2)(n2l)] (r1) {};
\node[] at ($(r0.north west) + (0,0.3)$) (name_l) {\tiny $l$};
\coordinate (h0) at ($(n2) + (1.5,0.3)$);
\draw[big edge] (n2) to[out=0,in=-90] (h0);
\draw[big edge] (n1) to[out=200,in=-90] (h0);
\draw[big edge] (name_l) to[out=-90,in=90] (h0);

    \end{scope}
    \begin{scope}[local bounding box=rhs, shift={($(lhs.south east) + (1,0)$)}, anchor=south west, scope anchor]
      
\node[big site, ] (s1r) {0};
\node[_BIG_copyInfo, right=1.00 of s1r, label={[inner sep=0.5, name=n3l]north:{\sf\tiny CopyInfo}}] (n3) {};
\node[big site, right=1.40 of n3,] (s0r) {1};
\node[_BIG_p, right=1.00 of s0r, label={[inner sep=0.5, name=n1l]north:{\sf\tiny }}] (n1) {};
\node[_BIG_fB, fit=(n3)(n3l)(s1r), label={[inner sep=0.5, name=n2l]north:{\sf\tiny FB}}] (n2) {};
\node[_BIG_l, fit=(n1)(n1l)(s0r), label={[inner sep=0.5, name=n0l]north:{\sf\tiny L("China")}}] (n0) {};
\node[big region, fit=(n0)(n0l)] (r0) {};
\node[big region, fit=(n2)(n2l)] (r1) {};
\node[] at ($(r0.north west) + (0,0.3)$) (name_l) {\tiny $l$};
\coordinate (h0) at ($(n2) + (1.5,0.3)$);
\draw[big edge] (n2) to[out=0,in=-90] (h0);
\draw[big edge] (n1) to[out=200,in=-90] (h0);
\draw[big edge] (name_l) to[out=-90,in=90] (h0);

    \end{scope}

    \node[xshift=0] at ($(lhs.east)!0.5!(rhs.west)$) {$\rrul$};
  \end{tikzpicture}      } 
     \caption{\rr{privacyViolation}: system rule that leads to a privacy violation by transferring the data to \keyword{FB} in \keyword{\mathtt{China}} .  } 
     \label{rr:violation}
\end{figure}

Defining system rules can inadvertently lead to privacy violations. As discussed in \cref{sec:Integrate}, data must not be transferred to \keyword{FB} in \keyword{\mathtt{China}} because the provided certification is invalid. However, modellers may define a system rule that allows data to be transferred to \keyword{FB}, as shown in \cref{rr:violation}. This violation occurs because the rule is defined without explicitly matching on \keyword{InvalidCert}. The PRISM model checker detects this violation as Property (3) in \cref{sec:verification} is not met (\ie the certification is invalid, but the transfer occurs regardless). By inspecting the generated transition system, it is straightforward to identify the rule introducing the violation, \ie the rule  generating a state in which simultaneously \pred{invalidCert} and \pred{dataTransfer} hold. To correct this rule, we redefine it as presented in \cref{rr:preventChina}.

Malformed initial states or reaction rules can introduce privacy violations. For instance, according to the defined sorting constraints, \keyword{L\mathtt{(Singapore)}} should contain only a single child of either sort \keyword{\mathtt{ctr}} \textit{or} \keyword{\mathtt{inctr}}. However, if a malformed initial state or rule assigns \keyword{L\mathtt{(Singapore)}} both sorts, this could lead to data being transferred to \keyword{L\mathtt{(Singapore)}} during the execution of the transition system, violating the intended semantics.

Another example that highlights the vital role of sorts in ensuring the correctness of the model is when \keyword{Cont} is linked to a port of sort \keyword{\mathtt{sc}} instead of \keyword{\mathtt{p}}. In this scenario, the process of checking the region cannot be performed, resulting in a bug in the model. Such an issue can lead to significant consequences, especially when data transfer is critical. To maintain the integrity of the model, \keyword{Cont} must be linked to ports of sort \keyword{\mathtt{p}} and \keyword{\mathtt{sy}}, as specified in the sorts.

\section{Related Work}
\label{sec:related_Work}
There is a lack of using formal methods to prove systems' compliance with the GDPR requirements for international data transfer. However, multiple works have been proposed to prove the systems' compliance with other GDPR notions. For example, Karami \etal~\cite{karami2022dpl} define data protection language (DPL) as an object-oriented language that handles the GDPR notions of purpose and storage limitation, the right to be forgotten, and providing/withdrawing consent. They use multiset rewrite rules to formally model the operational semantics of DPL. The resulting model is checked using the model checker Maude~\cite{meseguer_TwentyYearsOfRewritignLogic}, but the checking process was \textit{partially automated} as they also provide a pen-and-paper proof. Another tool that checks these notions is the Model-Based approach to Identify Privacy Violations in software requirements (MBIPV)~\cite{ye2023mbipv}. It is a \textit{fully automatic} tool that detects GDPR violations by converting UML class diagrams into a Kripke model. The NuSMV~\cite{nusmv} model checker is used to detect the violations in the defined Kripke model. 

Kammueller~\cite{kammueller2018formal} uses the theorem prover Isabelle (HOL) to prove the compliance of IoT healthcare systems with the GDPR requirements for users' right to access their data, their right to delete their data and the compliance of system functions with these rights. However, the proving process is not entirely automated.

Milner's $\pi$-calculus \cite{milner1999communicating} is extended \cite{DBLP:journals/lmcs/KouzapasP17} by defining a set of privacy calculus syntax. Later, the privacy calculus is improved to model the GDPR notion of providing and withdrawing consent~\cite{vanezi2019towards}. 
Another work based on $\pi$-calculus involves using multiparty session types to develop D´alogoP tool that proves the GDPR notion of processing purposes within systems~\cite{vanezi2020dialogop}.

All the mentioned approaches use formal techniques to evaluate the systems' compliance with specific aspects of the GDPR. However, they are limited in verifying the GDPR requirements for international data transfers because they need to support \textit{spatial properties} as bigraphs.~\cite{milner2009space}. 

Recent \textit{informal} approaches are proposed to prove the GDPR requirements for international transfer. Guamán \etal~\cite{guaman2021gdpr} define a method that systematically analyses the compliance of Android mobile apps with the GDPR notion of international data transfer. Later, the method is converted to an automated tool~\cite{guaman2023automated} but does not handle the certification mechanisms as our model does.
Pascual \etal~\cite{pascual2024hunter} provide \textit{Hunter}, an automated tool for tracking anycast communications protocol that routes the data to the nearest server regardless of location. The proposed approach mainly relies on the adequacy decision but does not cover other requirements for international data transfer, \eg the use of safeguards. Unlike our approach, which \textit{formally} proves the GDPR requirements for cross-border data transfers, these methods use \textit{informal} techniques, potentially resulting in a less rigorous analysis of these requirements.

\section{Conclusion}
\label{sec:discussion}

We provide a bigraphical framework that captures the GDPR requirements for cross-border data transfer. We encode these requirements using CTL and prove them by PRISM. We also define sorting schemes to ensure that our framework is well-formed.

Although we use only one example to define the framework, we believe it can capture different systems. The multi-perspectives approach enables us to model the GDPR requirements for international data transfers independently from specific systems. This supports the framework's ability to model several systems. Using parameters also allows various developers to model different countries and criteria.

Using our framework requires collaboration between software engineers with a background in bigraphs theory and privacy experts. The engineers build the model and explain the data flow to the privacy experts. The privacy experts advise on the certification criteria and the validity of the SCC, or they even amend the model if the regulations are updated. Organisations can use our framework to prove their adherence to the international transfer requirements.

In future, we aim to extend the framework to capture other safeguards, \eg Binding Corporate Rules (BCRs) and Codes of conduct. We also aim to model the case of transferring data based on the user's consent. The model need to be applied to more examples and real-world systems to show its generality. An automated tool is also needed to define and check sorting schemes automatically.

\subsection*{Acknowledgements}
This work is partially supported by an Amazon Research Award on Automated Reasoning.

\nocite{*}
\bibliographystyle{eptcs.bst}
\bibliography{doi.bib}

\newpage
\appendix

\section{System Rules}

We present some specific system rules for the WhatsApp example:
\label{app:systemRules}

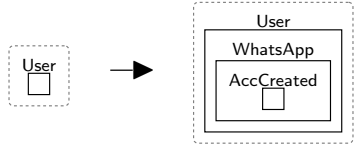
\begin{figure} [!ht]
    \centering
      \resizebox{0.3 \textwidth}{!}{
         \begin{tikzpicture}[
    ,
_BIG_whatsApp/.append style = {draw},
_BIG_user/.append style = {draw},
_BIG_accCreated/.append style = {draw}
    ]
    \begin{scope}[local bounding box=lhs, shift={(0,0)}]
      
\node[_BIG_user,  label={[inner sep=0.5, name=n0l]north:{\sf\tiny User}}] (n0) {};
\node[big region, fit=(n0)(n0l)] (r0) {};

    \end{scope}
    \begin{scope}[local bounding box=rhs, shift={($(lhs.south east) + (1.5,-0.45)$)}, anchor=south west, scope anchor]
      
\node[_BIG_accCreated,  label={[inner sep=0.5, name=n2l]north:{\sf\tiny AccCreated}}] (n2) {};
\node[_BIG_whatsApp, fit=(n2)(n2l), label={[inner sep=0.5, name=n1l]north:{\sf\tiny WhatsApp}}] (n1) {};
\node[_BIG_user, fit=(n1)(n1l), label={[inner sep=0.5, name=n0l]north:{\sf\tiny User}}] (n0) {};
\node[big region, fit=(n0)(n0l)] (r0) {};

    \end{scope}

    \node[xshift=0] at ($(lhs.east)!0.5!(rhs.west)$) {$\rrul$};
  \end{tikzpicture}      } 
     \caption{\rr{creatingAccount}: the system's user starts using WhatsApp to create an account. }   
     \label{rr:creatingAccount}
\end{figure}
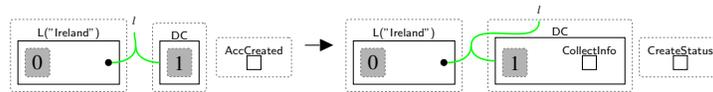
\begin{figure} [!ht]
    \centering
      \resizebox{0.6 \textwidth}{!}{

  \begin{tikzpicture}[
    ,
_BIG_p/.append style = {fill , circle ,  inner sep=1.3, black},
_BIG_l/.append style = {draw},
_BIG_dC/.append style = {draw},
_BIG_createStatus/.append style = {draw},
_BIG_collectInfo/.append style = {draw},
_BIG_accCreated/.append style = {draw}
    ]
    \begin{scope}[local bounding box=lhs, shift={(0,0)}]
      
\node[big site, ] (s1l) {0};
\node[_BIG_p, right=1.00 of s1l, label={[inner sep=0.5, name=n3l]north:{\sf\tiny }}] (n3) {};
\node[big site, right=1.00 of n3,] (s0l) {1};
\node[_BIG_accCreated, right=1.00 of s0l, label={[inner sep=0.5, name=n0l]north:{\sf\tiny AccCreated}}] (n0) {};
\node[_BIG_dC, fit=(s0l), label={[inner sep=0.5, name=n1l]north:{\sf\tiny DC}}] (n1) {};
\node[_BIG_l, fit=(n3)(n3l)(s1l), label={[inner sep=0.5, name=n2l]north:{\sf\tiny L("Ireland")}}] (n2) {};
\node[big region, fit=(n0)(n0l)] (r0) {};
\node[big region, fit=(n1)(n1l)] (r1) {};
\node[big region, fit=(n2)(n2l)] (r2) {};
\node[] at ($(r0.north west) + (-1.5,0.3)$) (name_l) {\tiny $l$};
\coordinate (h0) at ($(n3) + (0.5,0.3)$);
\draw[big edge] (n3) to[out=0,in=-90] (h0);
\draw[big edge] (n1) to[out=180,in=-90] (h0);
\draw[big edge] (name_l) to[out=-90,in=90] (h0);

    \end{scope}
    \begin{scope}[local bounding box=rhs, shift={($(lhs.south east) + (1,0)$)}, anchor=south west, scope anchor]
      
\node[big site, ] (s1r) {0};
\node[_BIG_p, right=1.00 of s1r, label={[inner sep=0.5, name=n4l]north:{\sf\tiny }}] (n4) {};
\node[big site, right=1.00 of n4,] (s0r) {1};
\node[_BIG_collectInfo, right=1.00 of s0r, label={[inner sep=0.5, name=n2l]north:{\sf\tiny CollectInfo}}] (n2) {};
\node[_BIG_createStatus, right=1.40 of n2, label={[inner sep=0.5, name=n0l]north:{\sf\tiny CreateStatus}}] (n0) {};
\node[_BIG_l, fit=(n4)(n4l)(s1r), label={[inner sep=0.5, name=n3l]north:{\sf\tiny L("Ireland")}}] (n3) {};
\node[_BIG_dC, fit=(n2)(n2l)(s0r), label={[inner sep=0.5, name=n1l]north:{\sf\tiny DC}}] (n1) {};
\node[big region, fit=(n0)(n0l)] (r0) {};
\node[big region, fit=(n1)(n1l)] (r1) {};
\node[big region, fit=(n3)(n3l)] (r2) {};
\node[] at ($(r0.north west) + (-2.0,0.3)$) (name_l) {\tiny $l$};
\coordinate (h0) at ($(n4) + (0.5,0.3)$);
\draw[big edge] (n4) to[out=0,in=-90] (h0);
\draw[big edge] (n1) to[out=180,in=-90] (h0);
\draw[big edge] (name_l) to[out=-90,in=90] (h0);

    \end{scope}

    \node[xshift=0] at ($(lhs.east)!0.5!(rhs.west)$) {$\rrul$};
  \end{tikzpicture}      } 
     \caption{\rr{creatingStatus}: the user creates a status, so the entity \keyword{AccCreated} is replaced with \keyword{CreateStatus}. We assume that \keyword{DC} collects information about the user's status, \protect\eg status content type, location information, information about the device used to post the status \protect\etc . We model that by generating the entity \keyword{CollectInfo}. We do not specify exactly what the collected data is, as our model focuses on checking the adequacy decision and the safeguards regardless of the type of transferred data. }   
     \label{rr:creatingStatus}
\end{figure}

\begin{figure} [!ht]
    \centering
      \resizebox{0.5 \textwidth}{!}{
         \begin{tikzpicture}[
    ,
_BIG_startTransfer/.append style = {draw},
_BIG_dC_Ir/.append style = {draw},
_BIG_collectInfo/.append style = {draw}
    ]
    \begin{scope}[local bounding box=lhs, shift={(0,0)}]
      
\node[big site, ] (s0l) {0};
\node[_BIG_collectInfo, right=1.00 of s0l, label={[inner sep=0.5, name=n1l]north:{\sf\tiny CollectInfo}}] (n1) {};
\node[_BIG_dC_Ir, fit=(n1)(n1l)(s0l), label={[inner sep=0.5, name=n0l]north:{\sf\tiny DC}}] (n0) {};
\node[big region, fit=(n0)(n0l)] (r0) {};
\node[] at ($(r0.north west) + (3.0,0.3)$) (name_l) {\tiny $l$};
\draw[big edge] (n0) to[out=0,in=-90] (name_l);

    \end{scope}
    \begin{scope}[local bounding box=rhs, shift={($(lhs.south east) + (1,0)$)}, anchor=south west, scope anchor]
      
\node[big site, ] (s0r) {0};
\node[_BIG_startTransfer, right=1.00 of s0r, label={[inner sep=0.5, name=n1l]north:{\sf\tiny StartTransfer}}] (n1) {};
\node[_BIG_dC_Ir, fit=(n1)(n1l)(s0r), label={[inner sep=0.5, name=n0l]north:{\sf\tiny DC}}] (n0) {};
\node[big region, fit=(n0)(n0l)] (r0) {};
\node[] at ($(r0.north west) + (3.0,0.3)$) (name_l) {\tiny $l$};
\draw[big edge] (n0) to[out=0,in=-90] (name_l);

    \end{scope}

    \node[xshift=0] at ($(lhs.east)!0.5!(rhs.west)$) {$\rrul$};
  \end{tikzpicture}      } 
     \caption{\rr{initialisingTrnas}: \keyword{DC} initialises the transferring process by replacing the entity \keyword{CollectInfo} with \keyword{StartTransfer} to start applying rule \rr{dcIrType} in \cref{rr:dcIrType}  }   
     \label{rr:initialisingTrnas}
\end{figure}
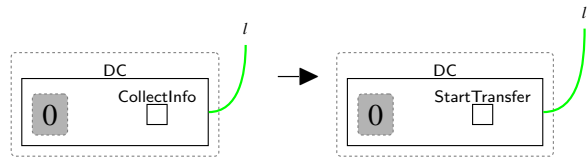

\begin{figure} [!ht]
    \centering
      \resizebox{0.7 \textwidth}{!}{

  \begin{tikzpicture}[
    ,
_BIG_withdReq/.append style = {draw},
_BIG_p'/.append style = {fill , circle ,  inner sep=1.3, blue},
_BIG_l/.append style = {draw},
_BIG_fB/.append style = {draw},
_BIG_compliantCert/.append style = {draw}
    ]
    \begin{scope}[local bounding box=lhs, shift={(0,0)}]
      
\node[big site, ] (s2l) {0};
\node[big site, right=1.00 of s2l,] (s0l) {1};
\node[_BIG_p', right=1.00 of s0l, label={[inner sep=0.5, name=n1l]north:{\sf\tiny }}] (n1) {};
\node[big site, right=1.00 of n1,] (s1l) {2};
\node[_BIG_fB, fit=(s2l), label={[inner sep=0.5, name=n3l]north:{\sf\tiny FB}}] (n3) {};
\node[_BIG_compliantCert, fit=(s1l), label={[inner sep=0.5, name=n2l]north:{\sf\tiny CompliantCert}}] (n2) {};
\node[_BIG_l, fit=(n2)(n2l)(n1)(n1l)(s0l), label={[inner sep=0.5, name=n0l]north:{\sf\tiny L("Dubai")}}] (n0) {};
\node[big region, fit=(n0)(n0l)] (r0) {};
\node[big region, fit=(n3)(n3l)] (r1) {};
\node[] at ($(r0.north west) + (4.2,0.3)$) (name_f) {\tiny $s$};
\node[left=3.0 of name_f] (name_l) {\tiny $l$};
\draw[big edge] (n2) to[out=0,in=-90] (name_f);
\coordinate (h0) at ($(n3) + (0.7,0.3)$);
\draw[big edge] (n3) to[out=0,in=-90] (h0);
\draw[big edge] (n1) to[out=90,in=-90] (h0);
\draw[big edge] (name_l) to[out=-90,in=90] (h0);

    \end{scope}
    \begin{scope}[local bounding box=rhs, shift={($(lhs.south east) + (1,0)$)}, anchor=south west, scope anchor]
      
\node[big site, ] (s2r) {0};
\node[_BIG_withdReq, right=1.00 of s2r, label={[inner sep=0.5, name=n4l]north:{\sf\tiny WithdReq}}] (n4) {};
\node[big site, right=1.40 of n4,] (s0r) {1};
\node[_BIG_p', right=1.00 of s0r, label={[inner sep=0.5, name=n1l]north:{\sf\tiny }}] (n1) {};
\node[big site, right=1.00 of n1,] (s1r) {2};
\node[_BIG_compliantCert, fit=(s1r), label={[inner sep=0.5, name=n2l]north:{\sf\tiny CompliantCert}}] (n2) {};
\node[_BIG_fB, fit=(n4)(n4l)(s2r), label={[inner sep=0.5, name=n3l]north:{\sf\tiny FB}}] (n3) {};
\node[_BIG_l, fit=(n2)(n2l)(n1)(n1l)(s0r), label={[inner sep=0.5, name=n0l]north:{\sf\tiny L("Dubai")}}] (n0) {};
\node[big region, fit=(n0)(n0l)] (r0) {};
\node[big region, fit=(n3)(n3l)] (r1) {};
\node[] at ($(r0.north west) + (4.2,0.3)$) (name_f) {\tiny $s$};
\node[left=3.0 of name_f] (name_l) {\tiny $l$};
\draw[big edge] (n2) to[out=0,in=-90] (name_f);
\coordinate (h0) at ($(n3) + (1.6,0.3)$);
\draw[big edge] (n3) to[out=0,in=-90] (h0);
\draw[big edge] (n1) to[out=90,in=-90] (h0);
\draw[big edge] (name_l) to[out=-90,in=90] (h0);

    \end{scope}

    \node[xshift=0] at ($(lhs.east)!0.5!(rhs.west)$) {$\rrul$};
  \end{tikzpicture}      } 
     \caption{\rr{ReqWithdCert}: \keyword{FB} in \keyword{\mathtt{Dubai}} asks to withdraw its
\keyword{certification}. The rule explicitly matches on the \keyword{CompliantCert} as we need to ensure that the entity requesting the withdrawal is using the certification, not the \keyword{SCCs}. }   
     \label{rr:ReqWithdCert}
\end{figure}
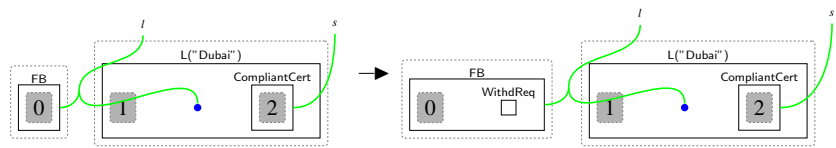

\end{document}